\documentclass[aps,prd,nofootinbib,twocolumn,floatfix]{revtex4-1}
\usepackage{graphicx}
\usepackage{dcolumn}
\usepackage{times,mathptm}
\usepackage{float}
\usepackage{color}
\usepackage{amsmath,amsfonts}
\usepackage{mathptmx}
\usepackage{mathrsfs}
\usepackage{bbm}
\usepackage{bm}
\usepackage{xfrac}

\DeclareMathOperator{\tr}{tr}

\newcommand{\beq}{\begin{eqnarray}}
\newcommand{\eeq}{\end{eqnarray}}

\newcommand{\D}{\mathcal{D}}
\renewcommand{\H}{\mathcal{H}}

\newcommand{\J}{\mathcal{J}}

\renewcommand{\d}{\delta}
\renewcommand{\l}{\lambda}

\renewcommand{\b}{\beta}
\renewcommand{\a}{\alpha}
\newcommand{\z}{\overline{z}}

\renewcommand{\o}{\omega}
\renewcommand{\tr}{\mbox{Tr}}

\newcommand{\tk}{\widetilde{k}}

\newcommand{\bx}{\mathbf{x}}

\newcommand{\m}{\mu}
\newcommand{\g}{\gamma}

\newcommand{\s}{\sigma}
\renewcommand{\k}{\kappa}
\renewcommand{\D}{\Delta}

\newcommand{\vA}{\vec{A}}

\newcommand{\N}{{\cal N}}

\renewcommand{\th}{\theta}

\newcommand{\oh}{\frac{1}{2}}

\newcommand{\dg}{\dagger}
\newcommand{\non}{\nonumber}

\newcommand{\rf}[1]{(\ref{#1})}
\newcommand{\ra}{\rightarrow}
\newcommand{\pa}{\partial}
\renewcommand{\vec}[1]{\bm #1}
\usepackage{ulem}

\begin{document}
\bibliographystyle{h-physrev5}

\title{Testing Proposals for the Yang-Mills Vacuum Wavefunctional \\  by
Measurement of the Vacuum}

\author{J. Greensite$^1$, H. Matevosyan$^2$, {\v S}. Olejn\'{\i}k$^3$, M. Quandt$^4$,
H. Reinhardt$^4$ and A. P.~Szczepaniak$^2$}
\affiliation{$^1$ Physics and Astronomy Dept., San Francisco State
University, San Francisco, CA~94132, USA \\
$^2$ Physics Department and Center for Exploration of Energy and Matter, Indiana University, Bloomington, IN 47403 USA \\
$^3$ Institute of Physics, Slovak Academy of Sciences, SK--845 11
Bratislava, Slovakia \\
$^4$ Institut f\"{u}r Theoretische Physik, T\"{u}bingen Universit\"{a}t, Auf der Morgenstelle 14,
T\"{u}bingen, D-72076 Germany}

\date{\today}
\begin{abstract}
   We review a method, suggested many years ago, to numerically measure the relative amplitudes of the true
Yang-Mills vacuum wavefunctional in a finite set of lattice-regulated field configurations.  The
technique is applied in 2+1 dimensions to sets of abelian plane wave configurations of varying amplitude and
wavelength, and sets of non-abelian constant configurations.  The results are compared to the predictions
of several proposed versions of the Yang-Mills vacuum wavefunctional that have appeared in the literature.  These
include (i) a suggestion in temporal gauge due to Greensite and Olejn\'{\i}k; (ii) the ``new variables" wavefunction put forward by Karabali, Kim, and Nair; (iii)  a hybrid proposal combining features of the temporal gauge and new variables
wavefunctionals; and (iv)  Coulomb gauge wavefunctionals developed by Reinhardt and co-workers, and by Szczepaniak and co-workers.  We find that wavefunctionals which simplify to a ``dimensional reduction" form at large scales, i.e.\ which have the form of a probability distribution for two-dimensional lattice gauge theory, when evaluated on long-wavelength configurations, have the optimal agreement with the data.
\end{abstract}

\maketitle

\section{\label{sec:intro}Introduction}

     Most of the key non-perturbative properties of non-abelian gauge theories, such as the static quark potential,
the chiral condensate, and the topological charge density, are actually properties of the vacuum of the quantized theory.  In the Hamiltonian formulation, the vacuum state is the ground state wavefunctional of the Hamiltonian operator, and all of the excited states of the theory, i.e.\ the mesons, baryons, and, in a pure gauge theory, the glueballs, are simply small excitations on top of that underlying ground state. For this reason, knowledge of the
Hamiltonian ground state wavefunctional could be essential in understanding the infrared 
properties of a non-abelian gauge theory.

    Proposals for the ground state of pure Yang-Mills theory go back over thirty years \cite{Greensite:1979yn,Halpern:1978ik}.  However, with only a few exceptions \cite{Mansfield:1993pd,Guo:1994vq,Kogan:1994wf,Samuel:1996bt,Haagensen:1995py}, very little work was done in this area after those initial efforts.  In recent years, however, there has been a modest revival of interest in this area, and a number of plausible suggestions for the vacuum state have been advanced.  These proposals will be described, along with their motivations, in the next section.  Briefly, there are  suggestions which have been put forward in temporal gauge \cite{Greensite:2007ij},  in Coulomb gauge \cite{Szczepaniak:2001rg,Szczepaniak:2003ve,Feuchter:2004mk,Reinhardt:2004mm,Feuchter:2007mq}  and, in 2+1 dimensions, in terms of  gauge-invariant ``new variables" \cite{Karabali:1998yq}.   Since these suggestions differ in various ways, it would be interesting to know which (if any) is the true vacuum state, or at least a reasonable approximation to the true vacuum state.

     In this article we will apply an old method \cite{Greensite:1987rg,Greensite:1988rr,Arisue:1992uq} for measuring, via lattice Monte Carlo simulations, the relative magnitudes of the true Yang-Mills wavefunctional in any given set of
lattice gauge field configurations.  The evaluations will be carried out for two types of lattice configurations:  non-abelian constant gauge fields of varying amplitudes, which are constant in space but noncommutative $[U_i,U_j] \ne 0$, and abelian plane waves of various amplitudes and wavelengths, which are abelian in the sense that $[U_i,U_j] = 0$.  The results are compared to the corresponding values obtained in each of the proposed vacuum wavefunctionals.  The method can be applied in any number of space-time dimensions, but here we will work exclusively in 2+1 dimensions, since the new variables proposal \cite{Karabali:1998yq} is formulated only in that case.

     In section \ref{proposals} below we will introduce and motivate each of the wavefunctionals to be tested.  Section
\ref{method} reviews the method for measuring the true vacuum wavefunctional, and section \ref{results} compares the results obtained by this method with the predictions of each of the proposed ground states.   Our conclusions are in section \ref{conclusions}, and some numerical details are found in the appendix.

\section{\label{proposals}Vacuum state proposals}

    The Yang-Mills Hamiltonian operator takes on its simplest form in temporal gauge, namely
\beq
H = \int d^{D}x \left\{ -\oh {\d^2 \over \d A^a_k(x)^2} + {1\over 4} F_{ij}^{a}(x)^2 \right\}
\label{basic-cont}
\eeq
in the continuum theory in $D+1$ dimensions, and
\beq
          H = {g^2 \over 2a} \sum_{l} E^a_l E^a_l + {1\over 2 g^2 a} \sum_p \mbox{Tr}[2 - U(p) - U^\dagger(p)]
\eeq
on the lattice, where the sums are over links $l$ and spatial plaquettes $p$, respectively.  Physical states in
temporal gauge must obey the Gauss law constraint ${D_k^{ab} E_k^b \Psi= 0}$, or more explicitly
\beq
 \Bigl( \d^{ac} \pa_k - g \epsilon^{abc} A^b_k \Bigr) {\d \over \d A^c_k }\Psi = 0  \; ,
\label{gauss}
\eeq
which implies that physical states must be invariant under infinitesimal gauge transformations.  The
Gauss law constraint in temporal gauge is a mixed blessing in the search for an approximate ground state.
On the one hand, gauge invariance can be seen as an aid in selecting a good ansatz for the vacuum state.
On the other hand, by severely limiting the choice, certain states which are perfectly acceptable in Coulomb gauge,
and which may be much more amenable to an analytical treatment, must be discarded in temporal gauge.  A very
important relation, for our purposes, is the equality of the vacuum wavefunctionals in temporal and Coulomb gauge
(see, e.g., ref.\ \cite{Greensite:2004ke}),
\beq
           \Psi_0^{Coul}[A] = \Psi_0^{temp}[A]
\label{equal}
\eeq
when evaluated on gauge fields satisfying the Coulomb gauge condition $\nabla \cdot A = 0$, and which also lie in
the first Gribov region.  Since our numerical method, to be described in the next section, will generate the relative
amplitudes of vacuum wavefunctionals in temporal gauge, in any finite set of gauge field configurations, we will be able to check proposals in Coulomb gauge by ensuring that the given set satisfies the Coulomb gauge condition, and lies within
the first Gribov horizon.

    The ground state wavefunctional is known in two limits:  the free-field $g^2=0$ limit, and also at strong lattice
couplings $g^2 \gg 1$.  In the free-field limit, in either Coulomb or temporal gauge,
\beq
\Psi_0[A] &=& \exp\left[-{1\over 4} \int d^{D}x d^{D}y ~ F_{ij}^a(x)
  \left({\d^{ab} \over \sqrt{-\nabla^2}} \right)_{xy} F_{ij}^b(y) \right]  \; ,
\non \\
\label{free}
\eeq
while in the strong-coupling limit, in $SU(N)$ gauge theory, it has been shown that \cite{Greensite:1979ha}
\beq
          \Psi_0[U] = \N \exp\left[ {N\over g^4 (N-1)} \sum_{P} \mbox{Tr}U(P) + \mbox{c.c.} \right]  \; ,
\label{strong}
\eeq
to leading order in $1/g^2$.   It was suggested long ago in ref.\ \cite{Greensite:1979yn}, by one of the present authors, that the Yang-Mills vacuum wavefunctional in 3+1 dimensions might have the form
\beq
              \Psi_0[A] \approx \Psi_0^{eff}[A] = \N \exp\left[-\oh \m \int d^3 x ~ \mbox{Tr}[F_{ij}^2] \right]  \; .
\label{YM-eff}
\eeq
when evaluated on sufficiently long-wavelength, slowly varying field configurations.  This wavefunctional has the property of {\it  dimensional reduction}:  If we write
\beq
           \Bigl| \Psi_0[A] \Bigr|^2 = \N e^{-R[A]}
\eeq
then $R[A]$ has the form of the Euclidean Yang-Mills action in one lower dimension (three dimensions, in this case).
It is clear that the strong-coupling vacuum state \rf{strong} does, in fact, have this property.

   The dimensional reduction vacuum \rf{YM-eff} in 3+1 dimensions is confining, i.e.
\beq
      W(C) &=& \langle \Psi_0 | \tr[U(C)] |\Psi_0 \rangle
\non 	\\
               &\sim& e^{-\mathrm{Area}(C)} \,
\eeq
if and only if Yang-Mills theory in three Euclidean dimensions has that property, where $U(C)$ is a Wilson loop holonomy around the planar, spacelike loop $C$.  Of course we have good reasons to believe that Yang-Mills theory is confining in three Euclidean dimensions.   It was noted by Halpern \cite{Halpern:1978ik} that a dimensional-reduction vacuum state in 2+1 dimensions {\it must} be confining, since Yang-Mills theory in two Euclidean dimensions is known to 
confine. Dimensional reduction was also suggested somewhat later, on rather different grounds, by Ambjorn, Olesen, and Peterson \cite{Ambjorn:1984mb,Ambjorn:1984dp}.  These authors were the first to make the connection between dimensional reduction and the property that has come to be known \cite{DelDebbio:1995gc} as Casimir scaling.  Strong evidence for Casimir scaling at intermediate distance scales was found in \cite{Bali:2000un}.

  On the other hand, the dimensional reduction wavefunctional cannot be correct as it stands,
because the short-distance structure is completely wrong.  For example, equal-time
two-point correlators in $D+1$ dimensions, at short distances, cannot be identical 
to short-distance two-point correlators in $D$ Euclidean dimensions; the singularity
structure in the approach to zero separation would be wrong.
In general one would expect that the vacuum state evaluated on short wavelength
configurations would agree with the perturbative ground state, whose zeroth order
approximation is given by \rf{free}.   

    There are other reasons, apart from short-distance singularity structure, that dimensional reduction cannot
be exact even for infrared physics.  Dimensional reduction from 2+1 to two Euclidean dimensions would imply a non-vanishing string tension, and perfect Casimir scaling, for any color group representation.  This cannot be right in 2+1 dimensions, because of color screening.\footnote{For this reason it is useful to consider $k$-string tensions, associated with quarks in completely antisymmetric representations, whose color charge cannot be screened to a lower dimensional representation by gluons.  The current evidence \cite{Bringoltz:2008nd} in 2+1 dimensions is that the leading corrections to the $N=\infty$ result are of order $1/N$ , as in Casimir scaling, rather than $1/N^2$, as in the competing Sine Law proposal.  For a recent discussion of $k$-string tensions in the context of the large-$N$ expansion, cf.\ \cite{Greensite:2011gg}. }
As argued in ref.\ \cite{Greensite:2007ij}, it is quite plausible that color screening is achieved by small corrections to the dimensional reduction form.  

    Another argument against exact dimensional reduction from 3+1 to three Euclidean dimensions was raised in refs.\ \cite{Burgio:2009xp,Quandt:2010yq}, which pointed out that this reduction would imply a match between the equal-time Coulomb gauge gluon propagator in $3+1$ dimensions, and the Landau gauge propagator in $D=3$ Euclidean dimensions.  It was shown in the same references that these propagators actually do agree quite well in a range of low and intermediate momenta around 1 GeV (a range which is relevant for phenomenology), but the equivalence cannot hold in the far infrared.

     For all of these reasons, a purely dimensional reduction vacuum wavefunctional is clearly inadequate.  Corrections are essential, and what is really required is an approximation to the vacuum state which holds at all distance scales.  There are now a number of proposals, which may or may not obtain the dimensional reduction form in some limit, but which do claim to approximate the ground state at all length scales.  These we will briefly review.

\subsection{Temporal gauge}

   It was suggested in ref.\ \cite{Greensite:2007ij} that the Yang-Mills ground state wavefunctional, in $D=2+1$ dimensions and in temporal gauge, is approximated by~\footnote{A factor of $g$ has been absorbed into the definition of the gauge field, so that $A_k$ has units of inverse length.  This accounts for the overall factor of $1/g^2$ in the exponent of the wavefunction.}

\beq
\Psi_{GO}[A] &=& \exp\left[-{1\over 2g^2}\int d^2x d^2y ~ B^a(x) \right.
\non \\
  & & \qquad \left. \times \left({1 \over \sqrt{-D^2 - \l_0 + m^2}} \right)^{ab}_{xy} B^b(y) \right]  \; ,
\label{GO}
\eeq
where $B^a=F^a_{12}$, $D^2$ is the covariant Laplacian, $\l_0$ is the lowest eigenvalue of $-D^2$, and $m^2$ is a parameter which vanishes as $g\ra 0$.  The motivation was to find the simplest possible gauge-invariant expression which would agree with the free-field \rf{free} and dimensional reduction \rf{YM-eff} wavefunctionals in the appropriate limits.  In support of this conjecture,
it was found that $\Psi_{GO}$
\begin{enumerate}
\item solves the Yang-Mills Schr\"{o}dinger equation in the strong-field, zero-mode limit;
\item confines if the mass parameter $m>0$, and that $m>0$ seems to be energetically preferred;
\item produces results for the mass gap, the Coulomb gauge ghost propagator, and the color Coulomb potential, which are in rather good agreement with results derived from standard lattice Monte Carlo simulations.
\end{enumerate}

    The subtraction of $\l_0$ is essential, and was introduced because $-D^2$ has a positive semi-definite spectrum, and in general the lowest eigenvalue tends to infinity for typical vacuum configurations in the continuum limit.  This fact is obvious perturbatively, and is confirmed numerically.  Without the subtraction (and this was the form originally suggested by Samuel \cite{Samuel:1996bt}), the kernel joining $B^a(x)$ and $B^b(y)$ in \rf{GO}
effectively vanishes in the continuum limit, and the corresponding string tension would be infinite.  In contrast, the spectrum of $-D^2 - \l_0$ is well-behaved, and not far from that of the free-field Laplacian operator $-\nabla^2$ \cite{Greensite:2007ij}.

     If one drops all components of the vector potential apart from the zero mode (analogous to the ``minisuperspace" approximation in quantum cosmology), then the Lagrangian and the Hamiltonian operators are simply
\beq
      L &=& {1\over 2g^2} \int d^{2}x ~ \Bigl[ \pa_{t} A_{k}\cdot \pa_{t} A_{k}
             - (A_{1} \times A_{2}) \cdot (A_{1} \times A_{2}) \Bigr]
\non \\
      &=& {1\over 2g^2} V \Bigl[ \pa_{t} A_{k}\cdot \pa_{t} A_{k}
             -  (A_{1} \times A_{2}) \cdot (A_{1} \times A_{2}) \Bigr]
\non \\
     H &=& -{g^2\over 2V} {\pa^{2} \over \pa A_{k}^{a} \pa A_{k}^{a} }
                 + {V\over 2g^2}  (A_{1} \times A_{2}) \cdot (A_{1} \times A_{2})  \; ,
\eeq
where $V$ is the volume of 2-space, and the cross-product and dot-product are defined with respect to $SU(2)$ color indices.  Solving for the ground state is a problem in quantum mechanics, rather than quantum field theory, and to
leading order in $1/V$ the solution is
\beq
           \Psi_{0} = \exp\left[- {V\over 2g^2} {(A_{1} \times A_{2}) \cdot
                      (A_{1} \times A_{2}) \over \sqrt{|A_{1}|^{2} + |A_{2}|^{2}}}  \right]  \; .
\label{zmode}
\eeq
Now in the region of parameter space where the zero mode is much larger than all other modes, the covariant Laplacian is approximated by
\beq
         (-D^{2})^{ab}_{xy} = \d^2(x-y) \Bigl[  (A_{1}^{2}+A_{2}^{2})\d^{ab} - A_{1}^{a}A_{1}^{b} -    A_{2}^{a}A_{2}^{b} \Bigr]
\non \\
\eeq
and $m^2$ is negligible. It is then found, after some algebra, that the proposed wavefunctional \rf{GO} reduces to the zero-mode solution \rf{zmode}.

    Dimensional reduction follows by expanding the $B$-field in eigenmodes $\phi_n^a$ of $-D^2$.  Then the part of the wavefunctional that depends only on the low-lying modes, with eigenvalues $\l_n - \l_0 \ll m^2$ has the form of the dimensional reduction wavefunctional \rf{YM-eff}, with $\m = 1/m$.  If we assume that the asymptotic string tension is due to the low-lying modes, then calculation of the string tension is simply an exercise in two-dimensional Yang-Mills theory, and the result is
\beq
            \s = {3\over 16} m g^2  \; ,
\eeq
If we turn this around, and write $m = 16 \s/(3 g^2)$, then we have a complete proposal for the vacuum wavefunctional, although the string tension must be supplied as an input.

    A method for obtaining equal time expectation values
\beq
            \langle Q \rangle = \int D A_k(x) Q[A] \Psi^2_{GO}
\eeq
by numerical simulation, with a suitable lattice regularization, was also introduced in \cite{Greensite:2007ij}, and applied
to calculate the mass gap.  The Coulomb gauge ghost propagator and color Coulomb potential were derived via numerical simulation of $\Psi^2_{GO}$ in \cite{Greensite:2010tm}, by the method of generating thermalized lattice configurations from the
$\Psi^2_{GO}$ distribution, and then transforming these configurations to Coulomb gauge.  The results, as already mentioned, were in very good agreement with those obtained from standard lattice Monte Carlo simulations. For details, we refer the reader to the cited references.

\subsection{New variables}

     While the temporal gauge ground state can be credited with some numerical success, it remains an educated guess, and requires the string tension as an input.  A more ambitious program in 2+1 dimensions, which aims to calculate both the Yang-Mills vacuum state and the string tension analytically, was initiated by Karabali, Kim, and Nair \cite{Karabali:1998yq}, and has been further developed by Karabali and Nair in a series of papers, cf. \cite{Karabali:2009rg} and references therein.

    The starting point in the Karabali, Kim, Nair (KKN) approach is temporal $A_0=0$ gauge, and the remaining two components of the $A$-field are combined into a complex field $A = (A_1+iA_2)/2$, related to a matrix-valued 
field $M$ via
\beq
\label{279}
               A =  - (\pa_z M)  M^{-1} ~~,~~  \overline{A} =   M^{\dg -1} \pa_{\z}  M^\dg  \; ,
\eeq
where $z=x_1-ix_2$, and $\z=x_1 + ix_2$ are the usual holomorphic variables in the complex plane.
The matrix-valued field $M$ takes values in the group $SL(2,C)$, and transforms covariantly, $M \ra G M$, under a gauge transformation $G$.  This field can be used to define
gauge-invariant field variables
\beq
           \H &=& M^\dg M
\non \\
            \J &=& {C_A \over \pi}  {\pa {\cal H}\over \pa z} {\cal H}^{-1}  \; ,
\eeq
where $C_A$ is the quadratic Casimir in the adjoint representation.
In terms of these gauge invariant variables, the
Hamiltonian becomes
\beq
H_{KKN} = T + V  \; ,
\eeq
where $T$ is derived from the $E^2$ term in the standard Hamiltonian
\beq
T &=& m \left( \int_u ~ \J^a(u) {\d \over \d \J^a(u)}+ \right.
\non \\
   & & \qquad \left. \int_{u,v} ~ \Omega_{ab}(u,v) {\d \over \d \J^a(u)} {\d \over \d \J^b(v)}\right)
\eeq
with
\beq
\Omega_{ab}(u,v) = {C_A \over \pi^2}{\d_{ab} \over (u-v)^2} - i f_{abc} {\J^c(v) \over \pi (u-v)}
\eeq
and ($\overline{\pa} \equiv \pa_{\z}$)
\beq
V = {1\over 2g^2} \int_x B^a(x) B^a(x) = {\pi \over mC_A} \int_z  \overline{\pa} \J^a \overline{\pa} \J^a
\eeq
and also
\beq
          m = {g^2 C_A \over 2 \pi}  \; .
\eeq
Inner products are evaluated with respect to
the integration measure
\beq
     \langle \Psi_1 | \Psi_2 \rangle = \int d\m(\H) ~ e^{2C_A S_{WZW}(\H)} \Psi^*_1(\H) \Psi_2(\H)  \; ,
\eeq
where $d\m(\H)$ is the  Haar measure, and $S_{WZW}$ is the Wess-Zumino-Witten action.

    Although the new field variable $\J$ is gauge invariant, the Hamiltonian $H_{KKN}$ is invariant under local
holomorphic transformations $h(z)$, under which $\J$ transforms like a connection
\beq
\J \ra h \J h^{-1} + {C_A \over \pi} \pa h h^{-1} \; ,
\eeq
and all physical states $\Psi[\J]$, in the new variables approach, must be invariant under this local transformation.
In this sense, the new variables approach trades the local gauge invariance constraint (the Gauss law) in temporal gauge
for invariance under local holomorphic transformations.

    Expressing the ground state as $\Psi_0[\J] = \N \exp(-R[\J])$, KKN find an expression for $R[\J]$ which is bilinear in
$\J$, namely
\beq
     \Psi_{KKN} &=& \N \exp\left[-{2\pi^2 \over g^2 C_A^2} \int d^2x d^2y ~ \overline{\pa} \J^a(x) \right.
\non \\
          & & \left.   \times  \left( {1 \over \sqrt{-\nabla^2 + m^2} + m} \right)_{xy}  \overline{\pa} \J^a(y) \right]
\non \\
     &=&  \N \exp\left[-{1\over 2g^2}\int d^2x d^2y ~ B^a(x) \right.
\non \\
    & & \left. \times \left( {1 \over \sqrt{-\nabla^2 + m^2} + m} \right)_{xy} B^a(y) \right]  \; ,
\label{KKN}
\eeq
where the second line is the new variables state converted back to usual variables.  KKN assume that the dimensional reduction form is obtained for long-wavelength configurations by simply dropping $-\nabla^2$ in the kernel, i.e.
\beq
     \Psi_{KKN} \ra \N \exp\left[-{1\over 2mg^2}\int d^2x  ~ B^a(x)  B^a(x) \right] \;  ,
\label{KKN-dimred}
\eeq
and then the string tension for a spacelike Wilson loop is obtained from solving Yang-Mills theory in two Euclidean
dimensions, with the result
\beq
         \s = {g^4 \over 8\pi} (N^2 - 1)  \; .
\eeq
Very remarkably, this value is within a few percent of the value found by Bringoltz and Teper \cite{Bringoltz:2006zg}
in lattice Monte Carlo simulations of the 2+1 dimensional theory, after careful extrapolation to the continuum
limit.\footnote{Recently some corrections to $\s$ have been calculated \cite{Karabali:2009rg}, and they are quite small.  At present it is not entirely clear {\it  why} the correction is so small, since there is no obvious small expansion parameter
in this approach,  and the corrections involve a sum of rather large (positive and negative) contributing terms, which for some reason nearly cancel.}

\subsection{\label{hybrid_prop}A hybrid wavefunctional}

   The problem with $\Psi_{KKN}$ is that, in terms of new variables, it is not holomorphic invariant, and in terms of the usual variables (second line of  \rf{KKN}) it is not gauge invariant.  Therefore $\Psi_{KKN}$, as it stands, is not a physical state.  Of course, KKN do not claim that $\Psi_{KKN}[\J]$ in eq.\ \rf{KKN} is exact, and presumably gauge and holomorphic invariance requires consideration of contributions
to $R[\J]$ involving higher powers of $\J$.  However, ignorance of the gauge/holomorphic-invariant wavefunctional calls
into question the assumed dimensional reduction form \rf{KKN-dimred}, which was required for the successful prediction of the string tension.  For example, suppose we assume that higher powers of $\J$ in the expansion of $R[\J]$ would have, as its main effect, the conversion of the ordinary Laplacian into a covariant Laplacian; i.e.\ in the usual variables
\beq
\Psi_0 &=& \N \exp\left[-{1\over 2g^2}\int d^2x d^2y ~ B^a(x) \right.
\non \\
 & & \left. \times  \left( {1 \over \sqrt{-D^2 + m^2} + m} \right)_{xy} B^a(y)\right] \; .
\eeq
In that case, for configurations which are non-abelian ($[A_x,A_y]\ne 0$) in general, dropping $-D^2$ is invalid even for configurations which vary very slowly compared to the length scale $1/g^2$, and indeed is invalid even for configurations which have no spatial variation whatever.  As we have remarked above, in connection with $\Psi_{GO}$, the covariant
operator $-D^2$  has a positive semi-definite spectrum, and for typical lattice configurations the lowest eigenvalue diverges in the continuum limit.  In that case, rather than replacing $-D^2$ by zero to obtain the dimensional reduction result, one should replace it by infinity!  This is obviously nonsense.

   Assuming that the KKN wavefunctional applies to abelian configurations ($[A_x,A_y] = 0$), the corresponding vacuum state for more general configurations is still a mystery; one can only guess what the gauge and holomorphic invariant completion of $\Psi_{KKN}$ might be. But the gauge-invariant completion is essential, if one is going to invoke dimensional reduction to compute the string tension.   At this stage there are an infinite number of possibilities, and the validity of the KKN prediction for the string tension depends on which of these possibilities is the correct one.  One possible approach is to retain $\Psi_{KKN}$ for abelian configurations, and ask for the simplest gauge-invariant generalization which would lead to the dimensional reduction form \rf{KKN-dimred}.  Then it is natural to merge features of $\Psi_{GO}$ and $\Psi_{KKN}$ into a conjectured ``hybrid" form for the ground-state wavefunctional
\beq
\Psi_{hybrid} &=& \N \exp\left[-{1\over 2g^2}\int d^2x d^2y ~ B^a(x) \right.
\non \\
   & & \times \left. \left( {1 \over \sqrt{-D^2 -\l_0+ m^2} + m} \right)^{ab}_{xy} B^b(y) \right]
\label{hybrid}
\eeq
which we will include in our numerical tests below.

   An alternative approach has been followed by Leigh, Minic, and Yelnikov (LMY) \cite{Leigh:2006vg}, who begin with the ansatz
\beq
\Psi_{LMY} =  \exp\left[-{\pi \over 2C_A m^2} \int d^2x d^2y ~ \overline{\pa} \J^a(x) K_{xy}(L)
                         \overline{\pa} \J^a(y) \right]  \; ,
\non \\
\label{LMY}
\eeq
where $L=-\D/m^2$, and $\D$ is the holomorphic-covariant Laplacian.  They then derive and solve a differential equation
for $K(L)$, where $L$ is treated as a number, rather than an operator, and by solving this equation they arrive at
\beq
           K(L) = {1\over \sqrt{L}} {J_2(4\sqrt{L}) \over J_1(4\sqrt{L}) }  \; .
\eeq
where $J_{1,2}$ are Bessel functions. By construction, the LMY proposal is a physical state.  If the infrared limit means $L\ra 0$, then $K \ra 1$, and $\Psi_0$ has the dimensional reduction form (\ref{KKN-dimred}), leading to the same prediction for the string tension.  Leigh et al.\ also obtain predictions for the glueball mass spectrum in 2+1 dimensions, which appear to be in good agreement with standard lattice Monte Carlo results.   The reservation in this case is that the LMY approach assumes a certain operator identity (eq.\ (56) of ref.\ \cite{Leigh:2006vg}) whose validity, in our opinion, is questionable.  It would nevertheless be interesting to test $\Psi_{LMY}$ numerically, but unfortunately it is not clear to us that the method we will use in this article could be easily applied to the LMY proposal.

\subsection{\label{TIU}Coulomb gauge}

In Coulomb gauge,  after  resolving Gauss' law, eq.~(\ref{gauss}), one obtains
the  Yang-Mills Hamiltonian \cite{Christ:1980ku} in terms of the transverse
components of the gluon field,    $\nabla\cdot\vec{A} = 0$,
\beq
\label{398-G1}
H &= & \frac{1}{2} \int d^D x \left( \J^{- 1} [A] \,{\Pi^a_i} \J [A] \,\Pi^a_i
+ B^a_i B^a_i \right) + H_c \\
H_c & = & \frac{g^2}{2} \int d^Dx d^Dy \,  \J^{- 1} [A] \,\rho^a (x)
\J [A] F^{ab} (x, y,[A]) \,\rho^b(y) \, , \nonumber
\eeq
where $\Pi^a (x) = \delta/i \delta A^a_i (x)$ is the canonical momentum
(electric field) operator and
\beq
\label{404}
\J [A] = \mathrm{Det} (- {D} \cdot {\nabla})
\eeq
is the Faddeev-Popov  (FP) determinant (this should not be confused with the variable $\J(x)$ in the KKN approach). Furthermore
\beq
\label{409-G3}
\rho^a (x) = - \epsilon^{abc} A^b_i \Pi^c_i
\eeq
is the color charge of the gluons and
\beq
\label{414-G4}
F^{ab}(x,y,[A])   = \Big[(- D\cdot \nabla)^{- 1}
\,(- \nabla^2)\, (- D\cdot {\nabla})^{- 1}\Big]_{x,a;y,b}
\eeq
is the so-called Coulomb kernel. The gauge fixed Hamiltonian
eq.~(\ref{398-G1}) is highly non-local due to the Coulomb kernel,
eq.~(\ref{414-G4}), and due to the FP determinant, eq.~(\ref{404}).
In addition, the latter occurs also in the functional integration measure of
the scalar product of Coulomb gauge wavefunctionals
\beq
\label{419-G5}
\langle \psi_1 | O | \psi_2 \rangle = \int D A \J [A] \,\psi^\ast_1 [A] \,O\, \psi_2 [A].
\eeq
Any normalizable state, expressed as a functional of the transverse gauge
field, is a physical state in Coulomb gauge.  This means in particular that a wavefunctional which is Gaussian in the gauge field may be a viable proposal for the ground state.  Unlike the GO and KKN/hybrid proposals, such a state cannot have the dimensional reduction property in general,
since that property calls for a wavefunctional which, on large scales, is Gaussian in the
field strengths rather than the gauge fields.  On the other hand, also unlike the other proposals,
the Gaussian wavefunctional is tractable analytically.

   Efforts in this direction were spearheaded by Szczepaniak and Swanson 
\cite{Szczepaniak:2000uf, Szczepaniak:2001rg}. They used a Coulomb gauge ground state wavefunctional of the form
\beq
\Psi [A] = \N \exp\left[-\oh\int \frac{d^Dk}{(2\pi)^D} \omega(k) A_i^a(k) A_i^a(-k) \right]. \label{IU0}
\eeq
The proposal was further developed in ref.~\cite{Szczepaniak:2003ve}, where the contribution from the
Faddeev-Popov determinant was included at one-loop order.
The field-independent function $\omega(k)$ was determined from a gap equation obtained by
minimizing the energy expectation value. The gap equation  depends on the so-called ghost dressing function $d(k)$, which is defined in terms of  the expectation value of the inverse Faddeev-Popov operator\footnote{As shown by Reinhardt \cite{Reinhardt:2008ek}, in Coulomb gauge the inverse ghost form factor $d^{- 1} (k)$ has the meaning of
the dielectric function of the Yang-Mills vacuum, and the horizon condition
\beq
\label{G52}
 d^{- 1} (0) = 0 
\eeq
 therefore implies that the Yang-Mills vacuum is a dual superconductor.}
   \beq
\int d^D x e^{ikx}  \langle  \Psi| \frac{g}{- (D \cdot {\nabla)}}  |\Psi \rangle _{x,a;0,b} =
\delta^{ab} \frac{d(k)}{k^2}
\eeq
and the Coulomb form factor, $f(k)$, defined by
\begin{equation}
\label{512}
 f(k) = \frac{\int d^D x e^{ikx} \langle \Psi| \left[\frac{{\nabla}^2}{ (- D \cdot {\nabla})  } \right]^2
  |\Psi \rangle_{x,a;0,b}}
{  \left[ \int d^Dx e^{ikx}  \langle  \Psi|  \frac{{\nabla}^2}{(- D \cdot {\nabla})} |\Psi
\rangle_{x,a;0,b}  \right]^2 }\, .
\end{equation}
In terms of $d(k)$ and $f(k)$ the expectation value of the Coulomb kernel in eq.~(\ref{414-G4}),
which determines the  Coulomb potential $V$, is given by
\begin{equation}
V(k) \equiv  \int d^D x \; e^{ik x } \langle \Psi | F^{ab}(x,0,[A]) |\Psi\rangle  = \delta^{ab}\frac{f(k)d^2(k)}{k^2}.
\label{gapeq}
\end{equation}
Finally, inclusion of the Faddeev-Popov determinant at one-loop order
introduces dependence on the function\footnote{For later use, we present all explicit expressions in $D = 2$ space
dimensions and for the color group $SU(N_C)$ \cite{Feuchter:2007mq}.}
($\hat k = k^i/|k|$)
\beq
\label{455-38-unreno}
\chi(k) &=& \frac{N_C}{2}  \int \frac{d^2 q}{(2 \pi)^2} \left[1 - (\hat{k} \cdot
\hat{q})^2\right]\,  {d(q) \,d(q-k) \over (q-k)^2}.
\eeq
which is related to the expectation value of ${\cal J}$. 
In ref.~\cite{Szczepaniak:2003ve} $\chi(k)$ (there denoted by $F(k)$)
was derived in context of the gap equation, while the explicit representation of ${\cal J}$ in terms of $\chi(k)$ was derived by Reinhardt and Feuchter in ref.~\cite{Reinhardt:2004mm} (cf. eq.~(\ref{430}) below). 

\noindent The set of coupled Schwinger-Dyson equations for $\chi(k),d(k),f(k)$ and $\omega(k)$
is UV divergent and  requires renormalization. In the variational approach this is achieved by
 adding  relevant and marginal counter-terms to the Hamiltonian and, if needed, renormalizing the functional  measure. The latter was obtained 
 in~\cite{Szczepaniak:2003ve} and reads 
  \begin{equation}
\label{455-38}
\chi (k) \rightarrow \chi(k,\mu)   =  I_\chi (k) - I_\chi (\mu) \; ,
\end{equation}
where $I_\chi(k)$ is given by the right hand side of eq.~(\ref{455-38-unreno}).
In \cite{Szczepaniak:2003ve} the  renormalization program was, however,  not  fully implemented. In particular a Hamiltonian counter-term  proportional to  $\int A \Pi$, which defines the $c_1$ renormalization constant (cf. eq.~(\ref{G49}) below),  was omitted  and thus only an approximate  low-energy solution could be obtained. It was found, however  to be  qualitatively consistent with the   results of ~\cite{Szczepaniak:2001rg}  that used the ${\cal J} = 1$ ($\chi(k)=0$)  approximation.  This hints that within the one-loop variational approach, contributions from the FP operator  may be accounted for by the gaussian wavefunctional itself, with an appropriate choice of the gaussian parameter $\omega(k)$. Such a possibility was rigorously demonstrated by Reinhardt and  Feuchter  ~\cite{Reinhardt:2004mm}   
(cf.\ eq.~(\ref{fullnewansatz}) below and the discussion that follows).

Inspired by the wavefunctional of a spinless particle in an s-state of a
spherical potential
  Feuchter and Reinhardt  in ~\cite{Feuchter:2004mk}
suggested to use
the ansatz
\begin{equation}
\label{544-XX}
\Psi [A] = \frac{\cal{N}}{\sqrt{\J [A]}} \exp \left[ - \frac{1}{2}
\int \frac{d^2 k}{(2 \pi)^D}\, \omega(k) A^a_i (k) A^a_i (- k) \right] \, ,
\end{equation}
which has
a number of technical advantages:
The factor of $\J[A]$ in the integration measure (eq.~(\ref{419-G5})) cancels
against $\J[A]^{-1}$ from the square of the wavefunction and thus drops out from the
calculation of equal-time vacuum expectation values.
As a consequence Wick's theorem can be applied directly, and in particular
$\omega(k)$ appearing in eq.~(\ref{544-XX}) is found to be directly related to the static gluon propagator
\beq
\langle A^a_i (k)   A^b_j (q) \rangle  =  (2\pi)^2 \delta^2(k+q)   \delta^{ab} \frac{  \delta_{ij}
 -\hat k_i\hat  k_j  }{2\omega(k) } .
\eeq
 In ref.~\cite{Reinhardt:2004mm} Reinhardt and Feuchter considered a general wavefunctional
 of the type
    \begin{equation}
\Psi_\alpha [A] = \frac{\mathscr{N}}{\J^\alpha[A]}\, \exp\left[ -
\frac{1}{2} \int \frac{d^2 k}{(2\pi)^D}\, A(-k)\,\omega_\alpha(k)\,A(k)\right] \, .
\label{fullnewansatz}
\end{equation}
In the one loop approximation they showed that the Faddeev-Popov
determinant, eq.~(\ref{404}), can be represented as
\beq
\label{430}
    \J[A] = \exp\left[-\int {d^2k \over (2\pi)^2}  \,A_i^a(-k) \,
    \chi(k)\,A_i^a(k) \right]
\eeq
where $\chi(k)$, thereafter referred to as the curvature, is given by
\begin{equation}
\label{577-XX}
\delta^{ab} \chi(k) = -\frac{1}{2} \int d^2x e^{ikx}
\left\langle \Psi_\alpha \left| \frac{\delta^2 \ln \cal{J}}{\delta A^a(x)
\delta A^b(0)} \right| \Psi_\alpha \right\rangle
\, ,
\end{equation}
  which, to the order of approximation considered, after renormalization, coincides with the one
  given in eq.~(\ref{455-38}). Combining eq.~(\ref{fullnewansatz}) and eq.~(\ref{430}) leads to
\begin{equation}
\Psi_\alpha[A] = \mathscr{N}\,\exp\left[- \frac{1}{2} \int
\frac{d^2k}{(2\pi)^2}\,A(-k)\,\big[\omega_\alpha(k) - 2 \alpha\,\chi(k)\big]\,A(k)\right]
\label{cgnewansatz}
\end{equation}
and establishes equivalence, at a one-loop level, 
between the ansatz of the Indiana group eq.~(\ref{IU0}),  which corresponds to $\alpha=0$, 
and that of the T\"uebingen group eq.~(\ref{544-XX}), corresponding to $\alpha=1/2$.\footnote{The value  of $\alpha$  does not matter in the one-loop approximation considered here.  It will, however, become relevant for calculations at higher loop order.}

However, using equivalent variational ans\"atze did not lead to the same results
for the correlation functions, $d(k)$, $f(k)$, $\chi(k)$, $\omega(k)$.  This is because the 
approaches of the Indiana and T\"ubingen groups  differ in {\it i)} 
the approximation scheme used to evaluate the expectation value of the Hamiltonian and 
 {\it ii)} the renormalization scheme. 
While the T\"ubingen group fully includes the Faddeev-Popov determinant to the
order considered, the Indiana group set $\J = 1$ throughout ref.~\cite{Szczepaniak:2001rg}
and neglected $\J$ in the Coulomb term in the numerical calculations of
ref.~\cite{Szczepaniak:2003ve}. (In the analytic calculation of
ref.~\cite{Szczepaniak:2003ve} $\J$ was, however,  fully included.)
Also, while the Indiana group considers the  one-loop corrections to the Coulomb form factor $f(k)$, the T\"ubingen group employs the $d(k)=1$ approximation in the equation for $f (k)$.

% In ref.~\cite{Szczepaniak:2003ve}  the renormalization program was not fully implemented. 
Ref.~\cite{Szczepaniak:2003ve}, in which the renormalization program was not fully implemented, missed a Hamiltonian counter-term proportional to $\int A \Pi$, which defines the $c_1$ renormalization constant (cf. eq. (\ref{G49}) below). The existence of this term was realized by Feuchter and  Reinhardt \cite{Feuchter:2004mk}, who carried out the complete renormalization program.
The $c_1$ counter-term missed in \cite{Szczepaniak:2003ve}
plays an important role  in determining the IR properties of the wavefunctional, 
as realized by Reinhardt and Epple \cite{Reinhardt:2007wh}, and will be crucial for the investigations given in the present paper.  Therefore throughout this paper we will use the fully renormalized  approach of the T\"ubingen group  
\cite{Feuchter:2004mk,Reinhardt:2007wh}.
      
 For later convenience we define
  \begin{equation}
 \overline \o(k)  \equiv \omega(k) - \chi(k),  \label{wf-TIU}
 \end{equation}
 where $\omega(k)$ corresponds to the wave functional in eq.~(\ref{544-XX}), 
 and write the wave functional of eq.~(\ref{544-XX}) in the form
 \begin{eqnarray}
 \Psi_{CG} [A] =
 \mathscr{N}\,\exp\left[- \frac{1}{2} \int
\frac{d^2k}{(2\pi)^2}A(-k) \overline \omega(k)  \,A(k)\right]. \nonumber \\
 \label{477-42}
 \end{eqnarray}
 The fully renormalized gap equation for $\omega$, which ultimately determines $\overline{\omega}$, reads ~\cite{Feuchter:2004mk,Reinhardt:2007wh}
\begin{eqnarray}
\o^2(k) = k^2 + \chi^2(k) + c_2 + \Delta I^{(2)}(k)  + 2 \chi(k)
\,[\Delta I^{(1)}(k) + c_1], \nonumber \\
\label{G49}
\end{eqnarray}
with
\beq
 \Delta I^{(n)}(k)  & =  &
 I^{(n)}(k) - I^{(n)}(0) \, ,
\non \\[2mm]
I^{(n)}(k) &=& \frac{N_C}{2}  \int \frac{d^2 q}{(2 \pi)^2} (\hat{k} \cdot
\hat{q})^2 \,V(q-k) { {\overline \o}^n(q) - {\overline \o}^n(k) \over \o(q) } \; ,
\nonumber \\
\eeq
and $V(k)$ given by eq.~(\ref{gapeq}). The gap equation, together with eq.~(\ref{455-38}) and the Schwinger-Dyson  equations for the ghost  form factor,
\begin{eqnarray}
\label{G45}
d^{-1}(k) &  = &  d^{-1}(\mu) - (I_d(k) - I_d(\mu) ),    \nonumber \\
I_d(k) & \equiv  & \frac{N_C}{2}  \int \frac{d^2 q}{(2 \pi)^2}\, \left[1 - (\hat{\k} \cdot
\hat{q})^2\right]\,{d(q-k) \over \o(q)\, (q-k)^2}
\end{eqnarray}
and Coulomb form factor,
\begin{eqnarray}
f(k) & =  & f(\mu) + ( I_f(k)  - I_f(\mu) ) \nonumber \\
I_f(k) & \equiv  & \frac{N_C}{2}  \int \frac{d^2 q}{(2 \pi)^2}\, \left[1 - (\hat{k} \cdot
\hat{q})^2\right]\,{f(q-k) d^2(q-k) \over \o(q)\, (q-k)^2}
\end{eqnarray}
form a closed set of coupled integral equations for $\chi,d,f$ and $\omega$.
In the gap equation (\ref{G49}), $c_1$ and $c_2$ are (finite) renormalization
constants. For the critical solution, where one imposes the horizon condition
 for the ghost dressing function, eq.~(\ref{G52}), both $\o(k)$ and $\chi(k)$ are infrared divergent,  which implies that the
 transverse gluon propagator vanishes at $k \ra 0$, while \cite{Reinhardt:2007wh}
\beq
\overline \o(0) \equiv \lim_{k \ra 0} (\o(k) - \chi(k)) = c_1.
\label{renormc1}
\eeq
So even when enforcing the horizon condition, the quantity $c_1 = \overline\o(0)$
is undetermined and may be taken to be either infrared finite or zero.
However, a perimeter law of the 't Hooft loop  requires $c_1 = 0$
and this value is also favoured by the variational
principle \cite{Reinhardt:2007wh}.  Furthermore, for $c_1 = 0$, in the IR limit $k \rightarrow 0$,
the wavefunctional eq.~(\ref{477-42}) becomes independent of the gluon zero mode
which agrees with the behavior of the exact vacuum wavefunctional in $1 + 1$ dimensions
\cite{Reinhardt:2008ij}, and corresponds to
the so-called ghost loop dominance in higher dimensions \cite{Zwanziger:2003de}.
But although there is strong evidence to
favor $c_1 = \overline\o(0) = 0$, our numerical studies in Section \ref{coulomb-TIU}
will also look at the case of a non-zero, but small, value for
$\overline\o(0)$. The renormalization parameter $c_2$, on the other hand,
has no influence on the IR or UV behavior of the solutions of the gap equation
(\ref{G49}). Only the mid momentum regime of $\omega (k)$ is weakly dependent
on $c_2$ \cite{Feuchter:2004mk}. Since we are mainly interested in the IR
properties we will put $c_2 = 0$ throughout this paper.

  The set of coupled integral equations can be solved analytically in the IR (for the critical solution)  using the power law ans\"atze \cite{Feuchter:2004mk,Schleifenbaum:2006bq} while the
 full numerical solutions of the above equations were given, for $D = 3$
space dimensions, in ~\cite{Feuchter:2004mk,Epple:2006hv,Epple:2007ut}.
For $D = 2$, the numerical solution was presented in ref.~\cite{Feuchter:2007mq} and it will be used in Section~\ref{coulomb-TIU} for comparison with lattice simulations.

      One criticism that can be leveled at the Coulomb gauge proposal is
     that it is not clear how it could ever
     lead to an area law falloff for spatial Wilson loops.
     In order to address this issue, a modified version of a Gaussian ansatz,  which
     incorporates monopole configurations, has been proposed by
     Matevosyan and Szczepaniak \cite{Szczepaniak:2010fe}. Furthermore, recently \cite{Campagnari:2010wc} Campagnari and Reinhardt
have developed a method which allows to use non-Gaussian wavefunctionals
in the variational approach. Specifically, a wavefunctional containing
vertices with up to four gluon fields was considered. Tests of these modified versions are, however, deferred to future
investigations.

\section{\label{method}The measurement method}

    We begin with the identity
\beq
           \Psi_0^2[U'_i(\bx)] &=& {1\over Z} \int DU ~ \left\{\prod_\bx \  \prod_{k=1}^2 \d[U_k(\bx,0)-U'_k(\bx)]
                                               \right\} e^{-S}
\non \\
\eeq
where, in the infinite volume limit, $\Psi_0$ is the ground state of the operator $H$, defined via the transfer matrix
$T=\exp[-Ha_t]$, with $a_t$ the lattice spacing in the time direction.  In the continuous time limit, $H$ is the Hamiltonian
of the lattice gauge theory.  Now consider a finite set of lattice configurations ${\cal U} \equiv \{U^{(m)}_k(\bx),m=1,2,...,M\}$ at a fixed time, and define
\beq
\widetilde{Z} = \sum_{m=1}^M  \int DU ~ \left\{\prod_\bx \  \prod_{k=1}^2 \d[U_k(\bx,0)-U^{(m)}_k(\bx)]
                                               \right\} e^{-S}
\eeq
This is the partition function of a statistical system in which the lattice configurations at time
$t=0$ are restricted to the set ${\cal U}$.  The rescaled wavefunctional
\beq
\lefteqn{\widetilde{\Psi}_0^2[U^{(n)}_i(\bx)]} & &
\non \\
   &=& { \Psi_0^2[U^{(n)}_i(\bx)] \over  \sum_{m=1}^M \Psi^2[U^{(m)}_i(\bx)] }
\non \\
 &=&    { \int DU ~ \left\{\prod_\bx \  \prod_{k=1}^2 \d[U_k(\bx,0)-U^{(n)}_k(\bx)] \right\} e^{-S}    \over
     \sum_{m=1}^M  \int DU ~ \left\{\prod_\bx \  \prod_{k=1}^2 \d[U_k(\bx,0)-U^{(m)}_k(\bx)]   \right\} e^{-S}  }
\non \\
\eeq
has the interpretation as the probability $P_n$ that, in this statistical system, a lattice configuration on the $t=0$ time-slice is equal to the $n$-th configuration $U^{(n)}_i(\bx) \in {\cal U}$ in the given set.

    The probability $P_n$ can be computed numerically by a modified lattice Monte Carlo simulation.  All links at $t\ne 0$
are updated in the usual way, which for the SU(2) gauge group with the Wilson action is a simple heat bath.  On the
$t=0$ plane, however, one of the $M$ configurations from the set ${\cal U}$ is selected at random, and then accepted
or rejected by the Metropolis algorithm.  Let $N_n$ represent the total number of times, in a given simulation, that the
$n$-th configuration in the set is selected by the Metropolis algorithm, with $N_{tot}$ the total number of updates of the $t=0$ plane.  Then
\beq
           P_n = \widetilde{\Psi}_0^2[U^{(n)}_i(\bx)] = \lim_{N_{tot} \ra \infty} {N_n \over N_{tot}} \; .
\eeq
Since $\widetilde{\Psi}_0[U^{(n)}]$ is simply a constant rescaling of $\Psi_0[U^{(n)}]$, it follows that
the relative amplitudes of the vacuum wavefunctional $\Psi_0$ in the set ${\cal U}$ are given by
\beq
           {\Psi^2_0[U^{(n)}] \over \Psi^2_0[U^{(m)}]} = \lim_{N_{tot} \ra \infty} {N_n \over N_m} \; .
\eeq

    Now suppose we have some theoretical proposal for the Yang-Mills vacuum wavefunctional
\beq
          \Psi_{theory}[U] = \N e^{-\oh R[U]} \; .
\eeq
If the proposal is correct, i.e.\ $\Psi_{theory}=\Psi_0$,  and we make a plot of
\beq
   -\log\left[{N_m \over N_{tot}}\right]   ~~ \mbox{vs.} ~~ R[U^{(m)}] \; ,
\eeq
then the data points should fall on a straight line, {\it  with slope equal to one}.

%    The method just described was introduced and applied some twenty years ago,
%in refs.\ \cite{Greensite:1987rg,Greensite:1988rr,Arisue:1992uq}.
%At that time there were no proposals for the Yang-Mills vacuum state at all distance scales, but only
%in the dimensional reduction limit at large scales, and computer simulations were limited to tests of the dimensional %reduction form on rather small lattices and lattice couplings $\b$.  Since we now have a number of concrete proposals %for the vacuum state at all scales, as outlined in the previous section, and more powerful
%computer resources, it is possible to greatly improve on these previous studies.

    The method just described was introduced and applied  in refs.\ 
\cite{Greensite:1987rg,Greensite:1988rr,Arisue:1992uq}.  In that early work, however, the simulations were carried out on small lattices and relatively small values of $\b=4/g^2$, while comparison to theory was limited to simple wavefunctionals, resembling \rf{strong}, inspired by the strong-coupling expansion.  It is now possible for us to greatly
improve on these previous studies.

    In this investigation we will consider sets of three different different types of configurations:
\begin{itemize}
\item Abelian plane waves with fixed wavelength $L$ and variable amplitude
\beq
            U^{(m)}_1(n_1,n_2) &=& \sqrt{1 - (a^{(m)}(n_2))^2} \mathbbm{1}_2 + i a^{(m)}(n_2)
            \s_3 \;
\non \\
            U^{(m)}_2(n_1,n_2) &=&    \mathbbm{1}_2
\non \\
             a^{(m)}(n_2) &=& {1 \over L} \sqrt{\a + \g m} \,\cos\left({2 \pi n_2 \over L}  \right) \; ,
\label{planewavesdef}
\eeq
where  $m=1,2,...,m_{max}$  with $L$ the lattice extension
and $\a,\g$ some constants.

\item Non-abelian constant configurations, variable amplitude:~\footnote{The factor of 20 in the definition of
$a^{(m)}$ is an arbitrary scaling of the parameters, which could of course be absorbed into $\a, \g$.}
\beq
            U^{(m)}_1(n_1,n_2) &=& \sqrt{1 - (a^{(m)})^2} \mathbbm{1}_2 + i a^{(m)} \s_1
\non \\
            U^{(m)}_2(n_1,n_2) &=&  \sqrt{1 - (a^{(m)})^2} \mathbbm{1}_2 + i a^{(m)} \s_2
\non \\
              a^{(m)} &=& \left[ {\a + \g m \over 20 L^2} \right]^{1/4} \; .
\label{nac1}
\eeq
\item Non-abelian constant configurations, fixed amplitude, variable ``non-abelianicity"  specified
by an angle $\theta_m$
\beq
      U^{(m)}_1(n_1,n_2) &=& \sqrt{1-\a^2}\mathbbm{1}_2  + i\a \s_1
\non \\
      U^{(m)}_2(n_1,n_2) &=& \sqrt{1-\a^2}\mathbbm{1}_2
\non \\
     & & \qquad + i\a(\cos(\theta_m) \s_1 + \sin(\theta_m) \s_2)
\non \\
      \theta_m  &=& \g (m-1) \pi  \; .
\label{nac2}
\eeq

\end{itemize}
\section{\label{results}Results}

     Since the measurement method in the previous section relies on a lattice regularization, we must apply this
regulator to the vacuum wavefunctionals under study.  Let us begin with $\Psi_{GO}$.  The proposal is that
\beq
          - \log[\Psi_{GO}^2[A]] = R_{GO}[A] + R_0 \; ,
\label{rescale}
\eeq
where $R_0=-\log(\N^2)$, and in the continuum
\beq
           R_{GO}[A] &=& {1\over g^2} \int d^2x \int d^2y ~ B^a(x)
\non \\
      & & \qquad  \times   \left[ {1\over \sqrt{-D^2 - \l_0 + m^2}}\right]^{ab}_{xy} B^b(y) \; .
\eeq
In the special case of abelian plane waves with $A^a_1(x) = A_1(x) \d^{a3}, ~ A_2^a(x) = 0$, we have the
simpler expression
\beq
          R_{GO}[A] &=& {1\over g^2}  \int d^2x \int d^2y ~ (\pa_2 A_1)_x
\non \\
     & & \qquad \times    \left[ {1\over \sqrt{-\nabla^2 + m^2}}\right]_{xy} (\pa_2 A_1)_y \; .
\eeq
The engineering dimension of the kernel, in 2+1 spacetime dimensions, is also inverse length.   We now latticize the theory and absorb
dimensions into a lattice spacing $a$, with
\beq
A_1(x) &\ra& {1\over a} A_{L1}(x) ~~,~~ \pa_2 \ra {1\over a} \pa_{L2} ~~,~~ \int d^2 x \ra a^2 \sum_x
\non \\
      g^2 &=& {g^2_L \over a} = {4\over \b a}  ~~,~~ m = {m_L \over a} \; ,
\label{latticize}
\eeq
where $\pa_L$ is the lattice finite difference operator, and all of the other subscript $L$ quantities are dimensionless.  All factors of $a$ cancel in $R[A]$, and the result is
\beq
       R_{GO}[A] = {\b \over 4} \sum_x \sum_y (\pa_{L2} A_{L1})_x  \left[ {1\over \sqrt{-\nabla_L^2 + m_L^2}}\right]_{xy} (\pa_{L2} A_{L1})_y \; .
\non \\
\eeq

\subsection{\label{planewaves}The GO and KKN wavefunctionals for abelian plane waves}

    Now we specialize to the lattice abelian plane wave configurations listed in the previous section (lattice sites are $x=(n_1,n_2)$)
\beq
            A_{L1}^{(j)}(n_2) {\s^3 \over 2}  &=& { U_1^{(j)}(n_1,n_2) -  U_1^{\dg(j)}(n_1,n_2)\over 2i}
\label{latcon}
\non \\
   U_2^{(j)}(n_1,n_2) &=& \mathbbm{1}_2
\non \\
            A_{L1}^{(j)}(n_2)  &=& {2\over L} \sqrt{\a +\g j} \cos\left({2\pi n_2\over L}\right)
\non \\
            \tk^2 &=& 2\Bigl(1-\cos\left({2\pi \over L}\right)\Bigr) \; .
\label{awave}
\eeq
Substituting these configurations into $R[A]$, the result is
\beq
           R_{GO}[U^{(j)}] = 2 (\a +\g j) \omega_{GO}(\tk^2) \; ,
\eeq
with
\beq
             \omega_{GO}(\tk^2) &=& {\b \over 4} { \tk^2 \over \sqrt{\tk^2 + m_L^2}}
\non \\
                                &=&  {1\over g^2}  {k^2 \over \sqrt{k^2 + m^2}} \; ,
\label{omGO}
\eeq
and where $k$ and $m$ are the momentum and the mass parameters in physical units, i.e.\ $k^2=\tk^2/a^2, m=m_L/a$.

    The same regularization applied to the KKN wavefunctional yields, for the abelian plane wave configurations,
\beq
           R_{KKN}[U^{(j)}] = 2 (\a +\g j) \omega_{KKN}(\tk^2) \; ,
\eeq
with
\beq
             \omega_{KKN}(\tk^2) &=& {\b \over 4} { \tk^2 \over \sqrt{\tk^2 + m_L^2} + m_L}
\non \\
                                &=&  {1\over g^2}  {k^2 \over \sqrt{k^2 + m^2} + m} \; .
\label{omKKN}
\eeq

\begin{figure}[t!]
\centerline{\scalebox{0.60}{\includegraphics{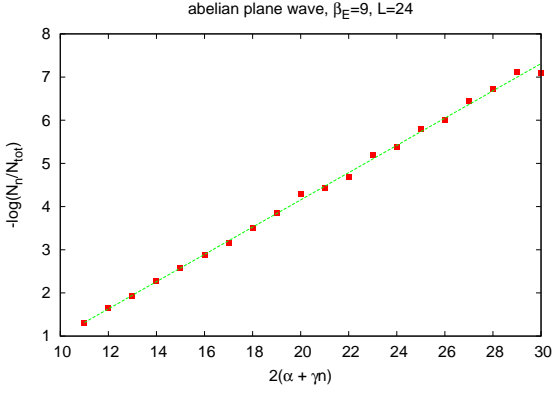}}}
\caption{A typical plot of the data for $-\log(N_n/N_{tot})$ at $\b_E=9$ and lattice extension $L=24$, vs. the factor
$2 (\a +\g n)$ associated with the amplitude of the $n$-th configuration.  The straight line is a best linear fit, and the
quantity $\omega_{MC}(\tk^2)$ is the slope of that line.}
\label{hugo1}
\end{figure}

    The theoretical values for $\omega(k^2)$ are to be compared against the data obtained from the numerical simulation.
For a given lattice coupling $\b_E$ of the Wilson action, at a given lattice size $L$ corresponding to a value of
$\tk^2$ given in eq.\ \rf{awave}, we obtain from the numerical simulation described in the previous section the values
\beq
             r_n = -\log\left({N_n \over N_{tot}}\right) \; .
\eeq
Then $\omega_{MC}(\tk^2)$ is obtained from a best linear fit of
\beq
              2 (\a +\g n) \omega_{MC}(\tk^2) + r_0
\label{fit}
\eeq
to the data points $\{r_n\}$.
Figure \ref{hugo1}
shows a typical plot of $r_n$ vs. $2(\a + \g n)$ at $\b_E=9$ and $L=24$; $\omega_{MC}(\tk^2)$ is the slope of the line (best linear fit) shown.  The values for $\a, \g$ used at each $\b_E$ and $L$ are listed in
Table \ref{Ag-ab} of the Appendix.

    The theoretical expressions for $\omega_{GO}(k^2)$ and $\omega_{KKN}(k^2)$ involve two dimensionful parameters $m$ and $g^2$.  Once these parameters are chosen, the results can be compared with the data obtained for $\omega_{MC}(\tk^2)$ on any
lattice, providing the dimensionless squared momentum $\tk^2$ on the lattice is converted into physical units
$k^2 = \tk^2/a^2$ using the lattice spacing $a$.  For a choice of lattice coupling $\b_E$, the lattice spacing in
physical units is given by
\beq
           a = \sqrt{\s_L \over \s}
\label{space}
\eeq
where $\s_L = \s_L(\b_E)$ is the $D=3$ dimensional string tension in lattice units, and $\s$ is the string tension
in physical units.  On grounds of tradition, we make the arbitrary choice $\s=(440~\mbox{MeV})^2$.

\begin{figure}[t!]
\centerline{\scalebox{0.72}{\includegraphics{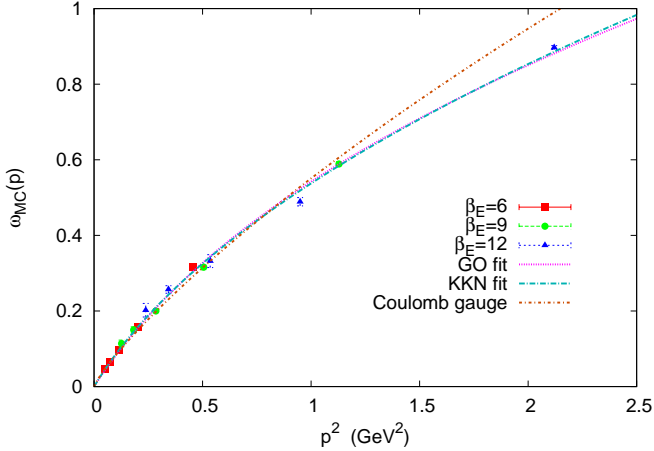}}}
\caption{Cumulative data for $\omega_{MC}$ vs.\ $p^2$ in physical units, on lattices of extensions
$L=16,24,32,40,48$, and Euclidean lattice couplings $\b_E=6,9,12$.  The curves labeled ``GO fit" and ``KKN fit" (there are actually two curves, difficult to distinguish from one another), are the theoretical values for $\omega_{GO}(p^2)$, and $\omega_{KKN}(p^2)$, using the parameters of $m$ and $g^2$ in Table I.  The line labeled ``Coulomb gauge" is obtained from the ansatz for the Coulomb gauge vacuum wavefunctional $\Psi_{CG} [A]$ (eq.~\ref{477-42}) as described in Section~\ref{coulomb-TIU}.}
\label{om}
\end{figure}

     Figure \ref{om} is a plot of $\omega_{MC}(k^2)$, extracted from a best fit of the data to eq.\ \rf{fit}.  Each data point is
obtained at a particular $\b_E=6,9$ or 12 on a given lattice of extension $L$, with $L=16,24,32,40$ or 48, and the wavelength of the plane wave on each lattice is the largest wavelength $\l=L$ available.   This plot
also displays the two theoretical curves
\beq
           \omega_{GO}(k^2) &=&   {1\over g^2}  {k^2 \over \sqrt{k^2 + m^2}}
\non \\
           \omega_{KKN}(k^2) &=&   {1\over g^2}  {k^2 \over \sqrt{k^2 + m^2} + m} \; ,
\eeq
with the parameters $g^2$ and $m$ obtained, for each curve, from a best fit to the data points.   Observe that in this range of momentum, the difference between the two fitting functions is essentially negligible, and in fact only becomes noticeable for $k^2 > 4$ GeV${}^2$.

     With the parameters obtained from the fit, we can use dimensional reduction (naively, in the KKN case, as explained in section \ref{hybrid_prop}) to compute the string tension, and compare it with our input value of (440 MeV)${}^2$.     Dimensional reduction gives
\beq
                 \sigma = m g^2 \times \left\{ \begin{array}{cl}
                         {3\over 16} & GO \cr
                                            &        \cr
                         {3\over 8}   & KKN \end{array} \right. \; .
\label{smg}
\eeq
The parameters $g^2,m$ from the best fit, and $\sqrt{\s}$ from obtained dimensional reduction,
in the GO and KKN cases are shown in Table I.  The values of $\sqrt{\s}$ should be compared with the given value of $\sqrt{\s} = 0.44$ GeV, which was used to set the lattice spacing at each $\b_E$.  The GO result is within 5\% of that value, and the KKN result is almost exactly right.

\begin{table}[h!]
\begin{center}
\begin{tabular}{|c|c|c|c|} \hline
   variant    &  $m$    &  $g^2$   &   $\sqrt{\s}$  from \\
                 &              &               &   diml red.   \\ \hline
   GO        &   0.771  &   1.465   &   0.460    \\
   KKN      &   0.420  &   1.237   &   0.441    \\  \hline
\end{tabular}
\end{center}
\caption{The parameters $m,g^2$ for the GO and KKN wavefunctionals,
determined from a best fit to the abelian plane wave data in Fig.\ \ref{om}, with $\sqrt{\s}$ derived from
dimensional reduction.  All values are in units of GeV. }
\end{table}

    The product of $m$ and $g^2$, in either the GO or KKN approach, determines the string tension $\s$ in either approach.   The dimensionless ratio $g^2/m$ is an output of the KKN approach, where it is predicted to be $\pi$.
If $m$ and $g^2$ are determined from a best fit to the data, then the actual ratio is $g^2/m=2.95$.  It is not clear, at this stage, whether this small discrepancy is significant, or should just be attributed to deviations from the continuum scaling due to a finite lattice spacing.

\subsection{ Tests of the Coulomb gauge wavefunctional}
\label{coulomb-TIU} 
To test the wavefunctional eq.~(\ref{477-42}), we first have to transfer
it to the lattice. We begin by rescaling the
gauge field $A_i \mapsto A_i / g$ so that a prefactor $g^{-2}$ appears in the
exponent of eq.~(\ref{477-42}), and $A_i(x)$ has engineering dimension of a mass.
With these conventions, the Fourier transformed kernel $\omega(k)$ and
curvature $\chi(k)$ also have dimensions of mass.

Next we latticize as in eq.~(\ref{latticize}) and rescale
the gauge field again to obtain the dimensionless field\footnote{Throughout
this section, we will denote dimensionless lattice objects with a caret.}
$\widehat{A}^c_k(\hat{x}) \equiv a\,A_k^c(a \hat{x})$.
For Coulomb gauge fixed connections, it is, in principle, important to use the so-called
midpoint rule when extracting the gauge fields from the lattice links $U_k$:
\begin{eqnarray}
U_k(\hat{x}) &=& a_k^0(\hat{x})\,\mathbbm{1} + i \,a_k^c(\hat{x})\,\sigma_c\nonumber \\[2mm]
\Longrightarrow\qquad\widehat{A}_k^c(\hat{x} + \sfrac{\hat{k}}{2})  &=&
-2 a_k^c(\hat{x}) \cdot \eta(a_k^0(\hat{x}))\,.
\label{midpoint}
\end{eqnarray}
As compared to simpler prescriptions such as eq.~(\ref{latcon}), we have
two modifications:
\begin{enumerate}
\item The shift in the argument on the lhs ensures
that the resulting lattice connection is exactly lattice transversal if the
link fields are,
\[
\nabla \cdot \widehat{A}(\hat{x}) = \sum_j \Big[\widehat{A}_j(\hat{x} + \hat{j}) -
\widehat{A}_j(\hat{x}) \Big] = 0\,.
\]
After Fourier transformation, the shift leads to a phase factor in the connection
which affects general observables but happens to drop out in the (quadratic)
exponent $R[A]$ tested here.
\item The $\eta$--correction in eq.~(\ref{midpoint}) comes from the $SU(2)$
algebra for parallel transporters over a finite distance $a$,
\[
\eta(t) = \frac{\arccos\,t}{\sqrt{1-t^2}} = 1 + \mathscr{O}(t^2)\,.
\]
It is only relevant for very strong fields far from the continuum limit. (In
our numerical studies, the correction never exceeded $5\%$.)
\end{enumerate}

\noindent
After Fourier transformation
\begin{equation}
\widehat{A}_i^c(k) = \sum_{\hat{x}}e^{- i k\hat{x}}\,\widehat{A}_i^c(\hat{x})\,,
\end{equation}
where $k_i = (2 \pi /L) \ell_i$ (with $-L/2 \le \ell_i < L/2$), a simple calculation leads to
the lattice version of the CG wavefunctional,
\begin{equation}
\begin{array}{r@{\,\,\,=\,\,\,}l}
R_{CG}[U] & \displaystyle\frac{1}{L^2}\sum_k\,\overline{\omega}(\overline{k})
\,\sum_{i=1}^2 \sum_{c=1}^3\,\left| \sum_{\hat{x}} e^{-i \hat{k} \hat{x}}\,
2 a_i^c(\hat{x})\,\eta(a_i^0(\hat{x}))\right|^2 + R_0
\\[5mm]
\displaystyle\overline{\omega}(\overline{k}) & \displaystyle
g^{-2}\,\big[ \omega(\overline{k}) - \chi(\overline{k}) \big] \,.
\end{array}
\label{res}
\end{equation}
Notice that the dimensionless momentum argument in the numerical continuum
solution of the gap equation is $k/g^2$, so that its lattice counterpart becomes
\begin{equation}
\overline{k}_i \equiv
\frac{2}{a\,g^2} \,\sin\left(\frac{\pi}{L}\,\ell_i\right)\,.
\end{equation}
To complete the lattice transcription, we only have to find an expression
for the function
\begin{equation}
h(\beta) \equiv a(\beta)\,g^2\,,
\label{hdef}
\end{equation}
where $\beta = 4 / (a g_0^2)$ is the usual lattice coupling for $SU(2)$ MC
simulations in $D=2+1$.  From high precision measurements of the
string tension in $D=2+1$ \cite{Teper:1998te}, the best fit in the scaling window
$\beta \in [3,12]$ is
\[
\hat{\sigma} = \sigma \,a^2 = \frac{b}{\beta^2} \,\left(1 + \frac{c}{\beta}
\right)
\]
with coefficients $b \approx 1.788$ and $c \approx 1.414$. From this,
\[
\hat{\sigma} = \sigma a^2 = \sigma \frac{16}{\beta^2 g_0^4} =
\frac{16\,\sigma}{\beta^2 g^4}\,\left[ 1 +
\mathscr{O}(\beta^{-1})\right] \stackrel{!}{=} \frac{b}{\beta^2}\,
\left( 1 + \frac{c}{\beta}\right)\,.
\]
From the leading terms of order $\mathscr{O}(\beta^{-2})$, we find
$b = 16\sigma / g^4$ and therefore
\begin{eqnarray}
h(\beta) &=& a \,g^2 = \sqrt{\sigma \,a^2}\frac{g^2}{\sqrt{\sigma}}
= \sqrt{\hat{\sigma}(\beta)} \frac{4}{\sqrt{b}} \nonumber \\[2mm]
&=&  \frac{4}{\beta}\,\sqrt{1 + \frac{c}{\beta}}\,,\qquad\qquad
c = 1.414\,.
\end{eqnarray}
This completes the lattice transformation of the Coulomb gauge wavefunctional.

Let us first look at the non-Abelian constant configurations (\ref{nac1}).
The corresponding lattice connection has the special colour structure
$A_i^c \sim \delta_i^c$, but is otherwise constant in space,
i.e.~Fourier transformation projects out the zero frequency contribution,
\[
\sum_{i=1}^2 \sum_{c=1}^3 |\widehat{A}_i^c(\vec{k})|^2 \sim  \delta_{\vec{k},\vec{0}}\,.
\]
The final result for the exponent in the wavefunctional
$\Psi_{CG}[A] \sim e^{- R_{CG}[A]/2}$ becomes, for non-Abelian constant configurations,
\begin{equation}
\begin{array}{r@{\,\,\,}c@{\,\,\,}l}
\displaystyle R_{CG}[U^{(m)}]
& = &\displaystyle 8 L^2 \,\arccos^2\left(\sqrt{1-(a^{(m)})^2}\right)\cdot
\overline{\omega}(0) + R_0 \\[4mm]
&\simeq &\displaystyle 8 L^2\,(a^{(m)})^2\cdot\overline{\omega}(0) + R_0 \,,
\end{array}
\label{1000}
\end{equation}
where the approximation in the second line comes from discarding the
$\eta$--correction in eq.~(\ref{midpoint}).

\begin{figure}[t]
\includegraphics[width=7cm]{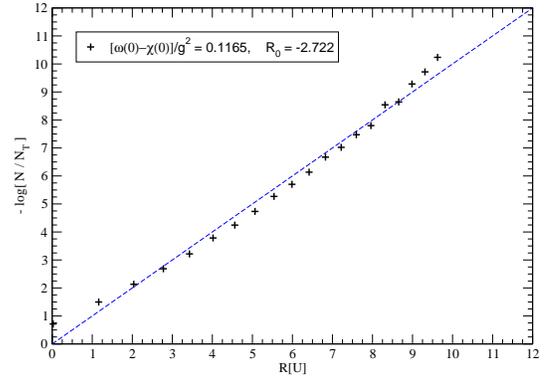}
\caption{The exponent $R$ from the variational approach eq.~(\ref{1000})
plotted against the lattice data for $-\ln \Psi^2$ for one set of
non-Abelian constant configurations, choosing $\overline{\omega} (0) = c_1$ as
fitting parameter ($c_1 = 0.1165$).}
\label{tuenac}
\end{figure}

From eq.~(\ref{renormc1}), the quantity $\overline{\omega} (0)$ is given by the
(finite) renormalization constant $c_1$ and, as already mentioned in sect.~\ref{TIU},
the energetically preferred value is $c_1 = 0$, which is also required for a
perimeter law in the 't Hooft loop \cite{Reinhardt:2007wh}. Obviously, with this
choice of renormalization constant $\overline{\omega} (0) = c_1 = 0$ the
 Coulomb gauge wavefunctional cannot account for the \textit{constant} non-Abelian
gauge field configurations. Whether this failure is important remains to be
seen. At least it does not necessarily imply that the Coulomb gauge wavefunctional
is a bad approximation to the true vacuum wavefunctional since constant
configurations form a set of measure zero in field space. One could give up the
preferred value $c_1 = 0$ and choose $\overline{\omega}(0) = c_1$ as a fitting
parameter, cf.~fig.\ref{tuenac}. This gives reasonable agreement with the
lattice data for one set of constant non-Abelian configurations but does not
cure the general problem. From the results presented in Sec.~\ref{kap4C} below,
it will become clear that constant non-Abelian gauge fields can only be accounted
for if we include quartic terms $\sim (\vA \times \vA)^2$ in the exponent of the
wavefunctional. The use of such non-Gaussian wavefunctionals in the variational
principle has recently become feasible \cite{Campagnari:2010wc}, but the
solution for the wavefunctional has not yet been determined explicitly up to
quartic terms in the exponent.

For these reasons, we will use the energetically favored value
$\overline{\omega} (0) = c_1 = 0$ in the following. We will now show that the
 Coulomb gauge wavefunctional does a good job
for Abelian plane waves of the type eq.~(\ref{planewavesdef}). In this case we have carried
out simulations at $\b=6$ on a fixed lattice volume of extension $L=24$, and varied the amplitude
of the plane waves, at given wavelength $L/M$, according to
\beq
            U^{(m)}_1(n_1,n_2) &=& \sqrt{1 - (a^{(m)}(n_2))^2} \mathbbm{1}_2 + i a^{(m)}(n_2)
            \s_3 \;
\non \\
            U^{(m)}_2(n_1,n_2) &=&    \mathbbm{1}_2
\non \\
             a^{(m)}(n_2) &=& {1 \over L} \sqrt{m \kappa_M} \,\cos\left({2 \pi n_2 M \over L} \right) \; ,
\label{planewavesdef1}
\eeq
where $m=1,...,m_{max}$, with $\kappa_M=1.4,0.45,0.17,0.09,0.036$ at $M=1,2,4,8,12$ respectively. The connection is Abelian, $A_i^c\sim \delta^{c3}$, with a harmonic spacetime dependence in the $y$-direction;
the corresponding wavenumber is proportional to the parameter $M$ in eq.~(\ref{planewavesdef1}).
After Fourier transformation the general result (\ref{res}) takes a
fairly complicated form
\begin{equation}
\begin{array}{r@{\,\,\,}c@{\,\,\,}l}
\displaystyle R_{CG}[U^{(m)}] &=& \displaystyle R_0 +
4 \sum_{n=-L/2 + 1}^{L/2}\overline{\omega}(p_n)
\Bigg|\sum_{r=0}^{L-1} \exp\left(- \frac{2\pi i}{L}\,n \,r \right) \\[5mm]
&&\displaystyle \times \mathrm{sgn}\,a^{(m)}(r)\cdot \arccos\sqrt{1 - (a^{(m)})^2(r)}
\Bigg|^2 \\[5mm]
\overline{p}_n &\equiv& \displaystyle \frac{2}{h(\beta)}\,
\sin\left( \frac{\pi}{L}\,n\right)  \; .
\end{array}
\label{2000}
\end{equation}
This can be simplified considerably, if the $\eta$--correction in the definition
of the connection, eq.~(\ref{midpoint}), is discarded.
Then the sums in eq.~(\ref{2000}) can be performed explicitly and we obtain
a much simpler expression
\begin{equation}
\begin{array}{r@{\,\,\,}c@{\,\,\,}l}
\displaystyle R_{CG}[U^{(m)}] &=& \displaystyle R_0 +
2 c_M\cdot m \kappa_M \cdot \overline{\omega}(\overline{p}_M)
\end{array} \; ,
\label{3000}
\end{equation}
where $c_M = 2$ for the highest frequency $M=L/2$ and $c_M = 1$ otherwise
for $L$ even ($L=24$ in this case). From eq.~(\ref{3000}), it is obvious that the plane wave
configuration tests the kernel $\overline{\omega} = \omega/g^2 - \chi/g^2$ exactly
at the lattice momentum $\overline{p}_M$ which corresponds to the frequency
of the plane wave.

\begin{figure}[t]
\includegraphics[width=7cm]{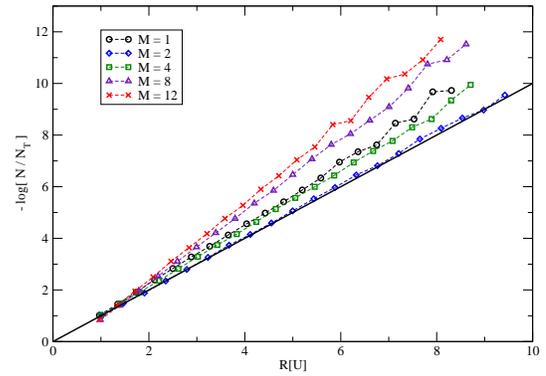}
\caption{The exponent $R_{CG}$ from the variational approach
eq.~(\ref{2000}) plotted against the lattice data for $-\ln \Psi^2$
for the plane wave configurations with wavenumber $M\in \{1,2,4,8,12\}$.
The lattice data was taken with lattice extension $L=24$ at $\beta = 6.0$.}
\label{tuewave}
\end{figure}

Figure \ref{tuewave} shows the result of the numerical evaluation of
eqs.~(\ref{2000}), (\ref{3000}) against the lattice MC data for Abelian plane
wave configurations of varying wavenumber and amplitude. As can be clearly
seen, the individual plane waves with fixed wavenumbers $M$ and varying amplitude
fall on a straight line, but the slope of that line differs from unity.
(We have chosen the solution $\overline{\omega}(k)$ of the variational problem
with the preferred renormalisation constant $c_1 = 0$.)
Morever, the slopes of the lines vary slightly with $M$, i.e.~effectively with
the momentum picked by the plane wave: For the smallest momentum $M=1$,
we find a slope of $1.19$, which decreases down to $1.02$ for $M=2$, and then
increases again up to $1.52$ for the largest momentum $M=12$ representable
on a $L=24$ lattice. If we relax the condition on the renormalisation constant
$c_1$ and take it as a free parameter, we observe that the spread in the slope
between the various wave numbers is increased, which is another hint that the
choice $c_1 = 0$ should be preferred.

Since the plane waves test the kernel $\omega(k)$ at varying momenta, we can
use a fit to the MC data as explained in the previous section to find a
numerical estimate $\omega_{MC}(k)$. In the Coulomb gauge wavefunctional,
this quantity corresponds to $\overline{\omega}(\overline{k}) =
g^{-2}\,\left(\omega(\overline{k}) - \chi(\overline{k})\right)$. After rescaling
to physical units (see eq.~(\ref{hdef}) and below), the result is plotted along with the
values obtained by numerical simulation,
$\omega_{MC}(k)$, in fig.~\ref{om}. It is evident that the variational solution
for $\overline{\omega}(k)$ fits the MC data very well, at least in the
infrared region for momenta up to $k \approx 1.3\,\mathrm{GeV}$. For larger
momenta, $\overline{\omega}(k)$ starts to deviate and becomes slightly
larger than the numerical estimate, but at most by a few percent within the
phenomenologically relevant mid-momentum regime. (For very large momenta
not plotted here, $\overline{\omega}(k)\sim k$ is exact by asymptotic freedom.)

%\begin{figure}[t!]
%\centerline{\scalebox{0.72}{\includegraphics{all-omega-and-TIU.eps}}}
%\caption{Open squares ($\Box$) represent the variational solutions for $\overline{\omega}(p)$ vs.\ $p^2$, obtained
%using the Coulomb gauge ansatz for the vacuum wavefunctional $\Psi_{CG}$.   These solutions are plotted together %with the Monte Carlo data $\omega_{MC}(p^2)$ 
%and the GO and KKN curves, that were already displayed in Fig.\ \ref{om}.}
%\label{om1}
%\end{figure}

\subsection{Non-abelian constant configurations: fixed amplitude, variable ``non-abelianicity"}\label{kap4C}

   For general non-abelian configurations we have, in a lattice regularization,
\beq
        R_{GO}[U^{(n)}] =  {\b \over 4} \sum_{x} \sum_{y} B^a(x) \left({1 \over \sqrt{-D^2 -\l_0 + m_L^2} }\right)_{xy}^{ab}
        B^b(y)
\non \\
\eeq
where
\beq
            B^a(x) = {1\over i} \mbox{Tr}[U(P_x) \sigma^a)]
\eeq
with $U(P_x)$ a product of links around a plaquette, starting with a link at site $x$. The lattice covariant Laplacian, in the adjoint representation, is given by
\beq
   (D^2)^{ab}_{xy} &=& \sum_{k=1}^2 \Bigl[ U^{ab}_k(x) \d_{y,x+\hat{k}}
         + U^{\dg ab}_k(x-\hat{k}) \d_{y,x-\hat{k}}  - 2 \d^{ab} \d_{xy} \Bigl]
\non \\
     U^{ab}_\m(x) &=& \oh \tr\Bigl[ \s^a U_k(x) \s^b U^\dg_k(x) \Bigr] \; .
\eeq
In terms of the parameters $g^2,m$ in the GO row of Table I, we use $\b=4/(g^2 a)$ and $m_L = m a$, where $a$ is the lattice spacing.  For comparison with the Monte Carlo data generated at the lattice coupling $\b_E$ of the Wilson action, we determine $a$ from eq.\ \rf{space}.   It is important to note that while we expect $\b / \b_E \ra 1$ in the continuum limit, this ratio need not be exactly equal to one at any finite $\b_E$.

   In the same way, the latticized ``hybrid" wavefunctional is
\beq
       \lefteqn{R_{hybrid}[U^{(n)}]}
\non \\
  &=&  {\b \over 4} \sum_{x} \sum_{y} B^{a}(x)
        \left({1 \over \sqrt{-D^2 - \l_0 + m_L^2} + m_L }\right)_{xy}^{ab} B^b(y) \; ,
\non \\
\eeq
with $\b,m_L$ determined using the parameters $g^2,m$ in the KKN row of Table I, and the lattice
spacing from eq.\ \rf{space}.

    We will consider first the configurations of eq.\ \rf{nac2}, with fixed amplitude and variable ``non-abelianicity" specified
by the $\th$ parameter.  If the amplitude is chosen small enough, then $-D^2 - \l_0$ is negligible compared to $m^2$,
and the kernel reduces to
\beq
\left({1 \over \sqrt{-D^2 - \l_0 + m^2}}\right)_{xy}^{ab} = {1\over m} \d_{xy} \d^{ab}
\eeq
for the GO wavefunctional, and
\beq
\left({1 \over \sqrt{-D^2 - \l_0 + m^2} + m}\right)_{xy}^{ab} = {1\over 2m} \d_{xy} \d^{ab}
\eeq
for the hybrid.  This is the dimensional reduction limit, and in either case, for the configurations \rf{nac2},
$R[U] \propto (A_1 \times A_2)^2$, or
\beq
            R_{GO,hybrid}[U^{(n)}] \propto \sin^2(\th_n)
\eeq
For the Coulomb gauge  wavefunctional, however, $R[U] \propto A_1^2 + A_2^2$, and hence, since the amplitudes of
$A_1$ and $A_2$ are fixed in the set \rf{nac2},
\beq
             R_{CG}[U^{(n)}] \propto \overline{\omega}(0)
\label{RCG0}
\eeq
independent of the angle $\th_n$.    If $\overline{\omega}(0)=0$, which seems optimal for agreement with the
plane wave data, then $R_{CG}$ would also be independent of the amplitude of the gauge fields.  However, it
is important to recall that the Coulomb gauge wavefunctional should not be evaluated outside the first Gribov horizon.
So even if $\omega(0)=0$, the restriction to the Gribov region amounts to a cutoff in the amplitude of
non-abelian constant configurations.

   The Monte Carlo simulation was carried out on a $12^3$ lattice at $\b_E=6$, with the $t=0$ configurations
chosen from
\beq
       U_1^{(n)} &=& \sqrt{1-\a^2} \mathbbm{1}_2 + i \a \s_1
\non \\
       U_2^{(n)}          &=& \sqrt{1-\a^2} \mathbbm{1}_2 + i \a (\cos(\theta_n) \s_1 + \sin(\theta_n) \s_2)
\eeq
with $\a=0.193$, and $\th_n = (n-1)\pi/38$.   By explicitly calculating numerically the lowest lying eigenvalues of
the lattice Faddeev-Popov operator, we have checked that these lattice configurations are all inside the first Gribov horizon.

\begin{figure}[htb]
\centerline{\scalebox{0.65}{\includegraphics{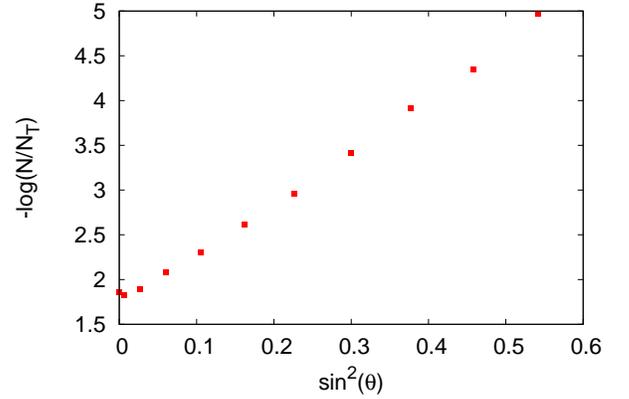}}}
\caption{Dependence of $-\log(N_n/N_T)$ on the "non-abelianicity" of the non-abelian constant configurations, determined by $\sin(\theta_n)$.}
\label{fig1}
\end{figure}

    In Fig.\ \ref{fig1} it can be seen that the logarithm of the wavefunctional is indeed proportional to $\sin^2(\theta)$, as one would expect from the GO and hybrid wavefunctionals in the dimensional reduction limit.  The data does not seem to be compatible, however, with the $\theta$-independence \rf{RCG0} of the CG wavefunctional \rf{477-42}.

\begin{figure}[htb]
\centerline{\scalebox{0.65}{\includegraphics{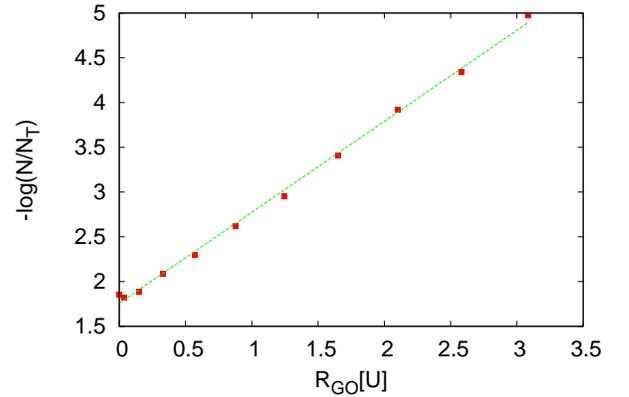}}}
\caption{Plot of $-\log(N_n/N_T)$ vs.\ $R_{GO}$ for the non-abelian constant configurations with
variable non-abelianicity.  The straight line fit has slope = 1.02.}
\label{fig2}
\end{figure}

   We recall that if $\Psi[U] = \exp[-\oh R(U)]$ is the true vacuum state, then the data points for $-\log(N_n/N_T)$ vs.\
$R[U^{n}]$ should fall on a straight line, with unit slope.
Plotting the data for $-\log(N_n/N_T)$ against $R_{GO}[U^{n}]$, as in Fig.\ \ref{fig2}, we find the slope obtained 
from a linear fit through the data is indeed close to unity.   In the GO case the slope is 1.02(6);  a similar analysis for the hybrid wavefunctional results in a slope of 1.12(7).  Some numerical details concerning the simulations are found in the Appendix.

\subsection{\label{D}Non-abelian constant configurations: variable amplitude, maximal ``non-abelianicity"}

  We now consider the non-abelian constant configurations of maximal ``non-abelianicity,", i.e.\ $\th=\pi/2$, which are the configurations of eq.\ \rf{nac1}, with index $m$ running from 1 to 20.
All Monte Carlo calculations were carried out on lattices of volume $32^3$ at $\b_E=6,9,12$, and the corresponding values of $\b, m_L$ at each $\b_E$ are given in Table \ref{beta}, where the values for the hybrid wavefunctional are taken to be the KKN values, since the hybrid reduces to the KKN form on abelian configurations.  The test of the GO and hybrid wavefunctionals is to see whether or not the data points for $-\log[N_n/N_{tot}]$, when plotted against $R[U^{(n)}]$, fall on a straight line whose slope is close to unity.

\begin{table}[htb]
\begin{center}
\begin{tabular}{|c|c|c|c|c|} \hline
   $\b_E$   &  $\b$ (GO)    &  $m_L$ (GO)  &   $\b$ (KKN)    &  $m_L$ (KKN)   \\ \hline
       6        &   4.73   &   0.445  &  5.60 & 0.242 \\
       9       &    7.43   &   0.283  &  8.80  & 0.154 \\
      12      &   10.19   &  0.207  &  12.07  &   0.113     \\  \hline
\end{tabular}
\end{center}
\caption{Values of $\b,m_L$ for the GO and KKN wavefunctionals at each $\b_E$, derived from the
$g^2,m$ parameters in Table I and the lattice spacings $a$, at $\b_E=6,9,12$.}
\label{beta}
\end{table}

\begin{figure}[thb]
\centerline{\scalebox{0.7}{\includegraphics{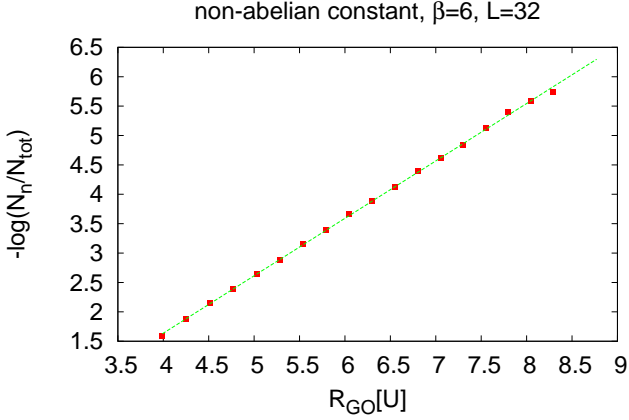}}}
\caption{Plot of $-\log(N_n/N_T)$ vs.\ $R_{GO}$ for non-abelian constant configurations, maximal non-abeliancity,
at $\b_E=6,~L=32,~\a=2,~\g=0.15$  In this case the straight line fit has a slope = 0.98.}
\label{b6add2}
\end{figure}

   An example of the $-\log[N_n/N_{tot}]$ vs.\ $R_{GO}[U^{(n)}]$ data at $\b_E=6$ is shown in Fig.\ \ref{b6add2}, for the
choice $\a=2, \g=0.15$.  Although the data is nicely fit by a straight line which has a slope close to unity, this fact must be interpreted with caution because, since the number $N_n$ falls off exponentially with $R_{GO}[U^{(n)}]$, the range of $R$ must necessarily be kept small; typically $\D R \approx 4-5$.  This {\it  could} mean that the tendency of the data to lie on a straight line
is misleading, and perhaps we are simply looking at the tangent of a curve.  It is therefore necessary to extract the slope
of the straight line over small intervals centered around points over a wide range of $R$. The question is whether those slopes are constant, in which case the linearity hypothesis is verified, or whether they vary significantly as $R$ increases.  This is the motivation to calculate $-\log[N_n/N_{tot}]$ in sets of twenty configurations, using different values of the parameters $(\a,\g)$ in each set.  The parameters we have used are shown in Table \ref{Ag-nab} of the Appendix.

     Figure \ref{slope}  is a plot of the slope vs. $R$ at $\b_E=6,9,12$, where the value of $R$ at each data point is
the midpoint of the range in which the slope was computed.  Things are not perfect; there is some slight variation in the
slope with $R$, there is a little variation with $\b$, and the values of the slope are not exactly one (they seem to be closer to $1.1$ at the large $R$ values).  On the other hand, we have made no claim that the GO wavefunctional is exact, nor is asymptotic scaling exact at these lattice couplings.  The point is that scaling is not bad, and the slopes are fairly close to unity over a large range of $R$, using $g^2,m$ values that were extracted from fits to a completely different type of lattice configuration (i.e.\ abelian plane waves).

\begin{figure}[t!]
\centerline{\scalebox{0.65}{\includegraphics{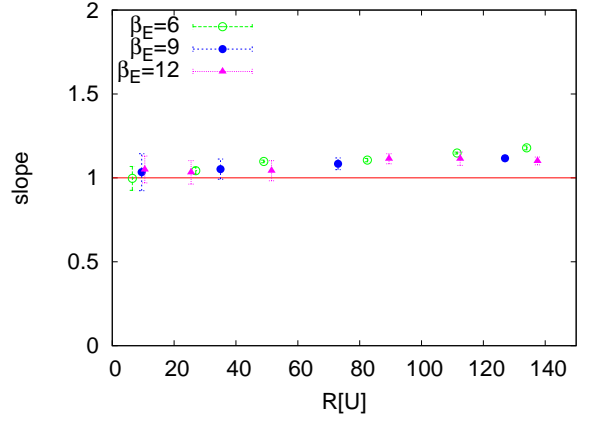}}}
\caption{Slopes for the GO wavefunctional vs. $R$, at $\b_E=6,9,12$ and $L=32$, using the values of
$g^2,m$ derived from the abelian plane wave fit.}
\label{slope}
\end{figure}

\begin{figure}[t!]
\centerline{\scalebox{0.65}{\includegraphics{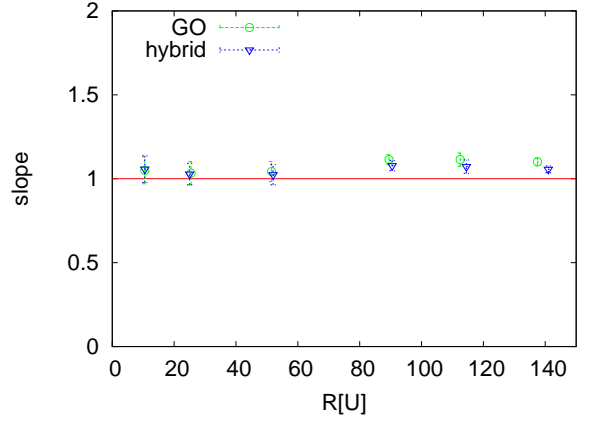}}}
\caption{$\b_E$=12 calculation, for both types of wavefunctionals.}
\label{compare}
\end{figure}

   Results for the hybrid wavefunctional  turn out to be quite close to those of the GO wavefunctional.
The values for $\b_E=12$, for both types of wavefunctionals, are shown in Fig.\ \ref{compare},
with similar agreement at the two other $\b_E$ values.

\subsection{The ghost propagator and the Coulomb potential}

    Because of the equality \rf{equal} of the vacuum wavefunctionals in temporal and Coulomb gauges, when
evaluated on transverse ($\nabla \cdot A = 0$) gauge fields,  equal-time expectation values in Coulomb gauge
can be derived from
\beq
        \langle Q \rangle = \int DA ~ Q[A] \d(\nabla \cdot A) \J[A] \Psi^2_0[A] \; ,
\label{cg-int}
\eeq
and we may use for $\Psi_0$ either of the temporal gauge proposals, $\Psi_{GO},~\Psi_{hybrid}$, or the Coulomb gauge proposal $\Psi_{CG}$  to calculate such objects as the ghost propagator
\beq
           G(R) = \left\langle \left( -{ 1\over \nabla \cdot D[A]} \right)^{aa}_{xy} \right\rangle_{|x-y|=R}
\label{gcoul}
\eeq
and the color Coulomb potential~\footnote{More precisely, for color charges in some representation $r$,
the Coulombic potential energy is obtained by multiplying $V_c(R)$ by the quadratic Casimir $C_r$, and dividing by
the dimension of the adjoint representation.}
\beq
          V_c(R) = - \left\langle \left(  {1\over \nabla \cdot D} (-\nabla^2) {1\over \nabla \cdot D}
                   \right)^{aa}_{xy} \right\rangle_{|x-y|=R} \; .
\label{vcoul}
\eeq
In eq.\ \rf{cg-int} there is an implicit restriction of the integration domain to the Gribov region. For the Coulomb gauge wavefunctional $\Psi_{CG}[A]$ the ghost propagator and the Coulomb potential are presented in ~\cite{Feuchter:2007mq}.

    In an ordinary Monte Carlo (MC) simulation, Coulomb gauge expectation values are obtained by first generating lattice
configurations with the usual probability distribution $\exp[-S]/Z$, where $S$ is the standard lattice action, transforming those configurations to Coulomb gauge, and evaluating the observable $Q$ in the ensemble of transformed configurations.  In principle the same strategy applies to evaluating the right hand side of \rf{cg-int} numerically; the problem in that case is to generate configurations with the probability distribution $\Psi^2[U]$, and this problem was solved, for the $\Psi_{GO}$ proposal, in ref.\ \cite{Greensite:2007ij}.  The simulation method developed in
\cite{Greensite:2007ij} is also applicable (although it has not been applied until now) to the hybrid proposal.  The lattice ghost propagator and Coulomb potential were calculated numerically from $\Psi_{GO}$, and compared to the corresponding results in ordinary lattice Monte Carlo, in ref.\ \cite{Greensite:2010tm}.  In that work, however, the authors chose $\b=\b_E$ and $m_L=4 \b \s_L/3$.  In the present article the philosophy has changed somewhat.  We have two parameters with dimensions of mass, $g^2$ and $m$,
and a scale set (arbitrarily) by taking $\sqrt{\s}=440$ MeV.  Then $g^2, m$ are chosen to give a best fit to the abelian plane wave data in Fig.\ \ref{om}.  To compare wavefunctional results with standard Monte Carlo results we determine the lattice spacing $a$, at each $\b_E$, from $\sqrt{\s_L/\s}$, and then $\b=4/(g^2 a)$ and $m_L=ma$
are the corresponding dimensionless parameters to use in the latticized wavefunctional $\Psi_{GO}$ or $\Psi_{hybrid}$.
With the new procedure we have $\b\ne \b_E$, and the obvious question is whether this fact will tend to destroy the agreement that was found previously, in
\cite{Greensite:2010tm}, between ghost propagators and Coulomb potentials derived from simulation of $\Psi^2_{GO}$, and the corresponding quantities found in ordinary lattice Monte Carlo simulations.  We would also like to calculate the Coulomb gauge ghost propagator and Coulomb potential for the hybrid wavefunctional proposal.  \\

\begin{figure}[t!]
\centerline{\scalebox{0.70}{\includegraphics{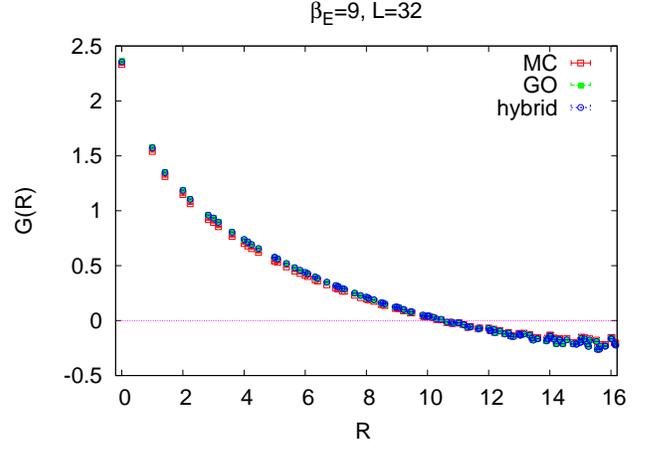}}}
\caption{The ghost propagator derived from standard Monte Carlo (MC) simulation at $\b_E=9$, and
the same quantity calculated by simulation of the GO and hybrid wavefunctionals, by the technique
described in ref.\ \cite{Greensite:2007ij}.}
\label{GOghost}
\end{figure}

\begin{figure*}
  \begin{center}
    \begin{tabular}{cc}
      \resizebox{70mm}{!}{\includegraphics{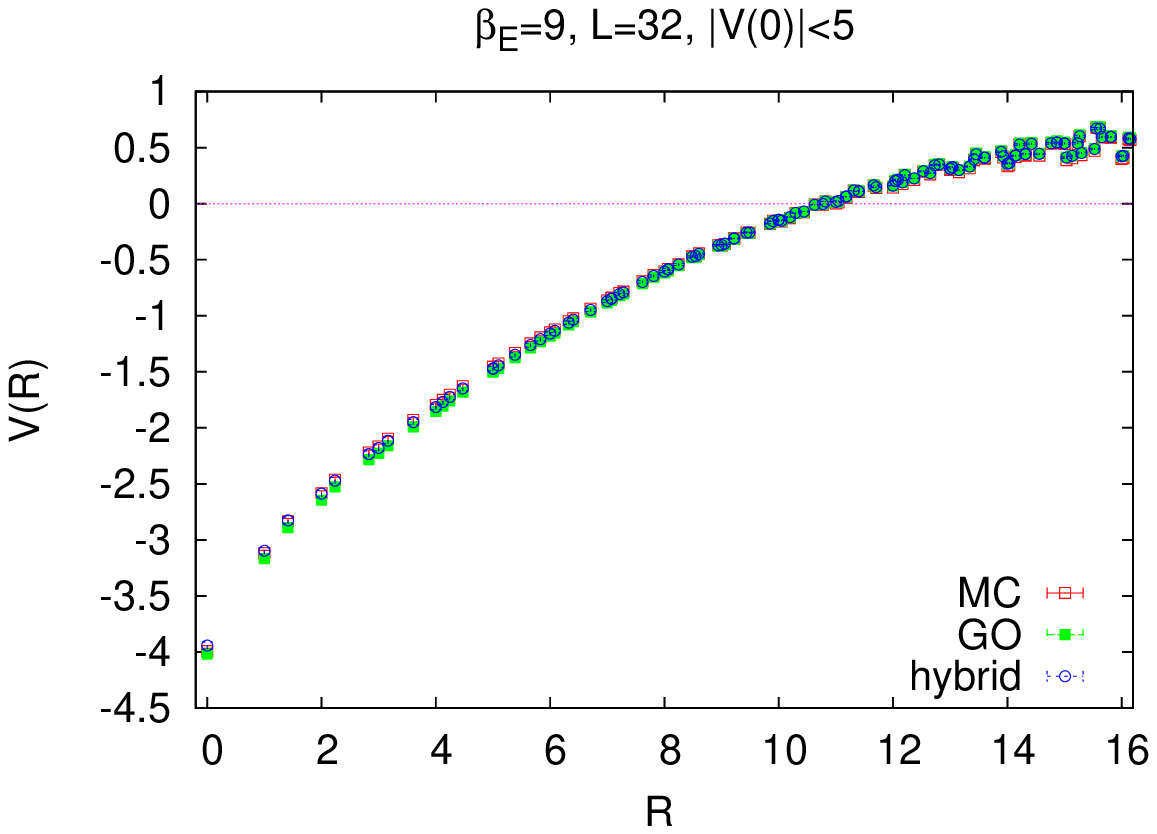}} &
      \resizebox{70mm}{!}{\includegraphics{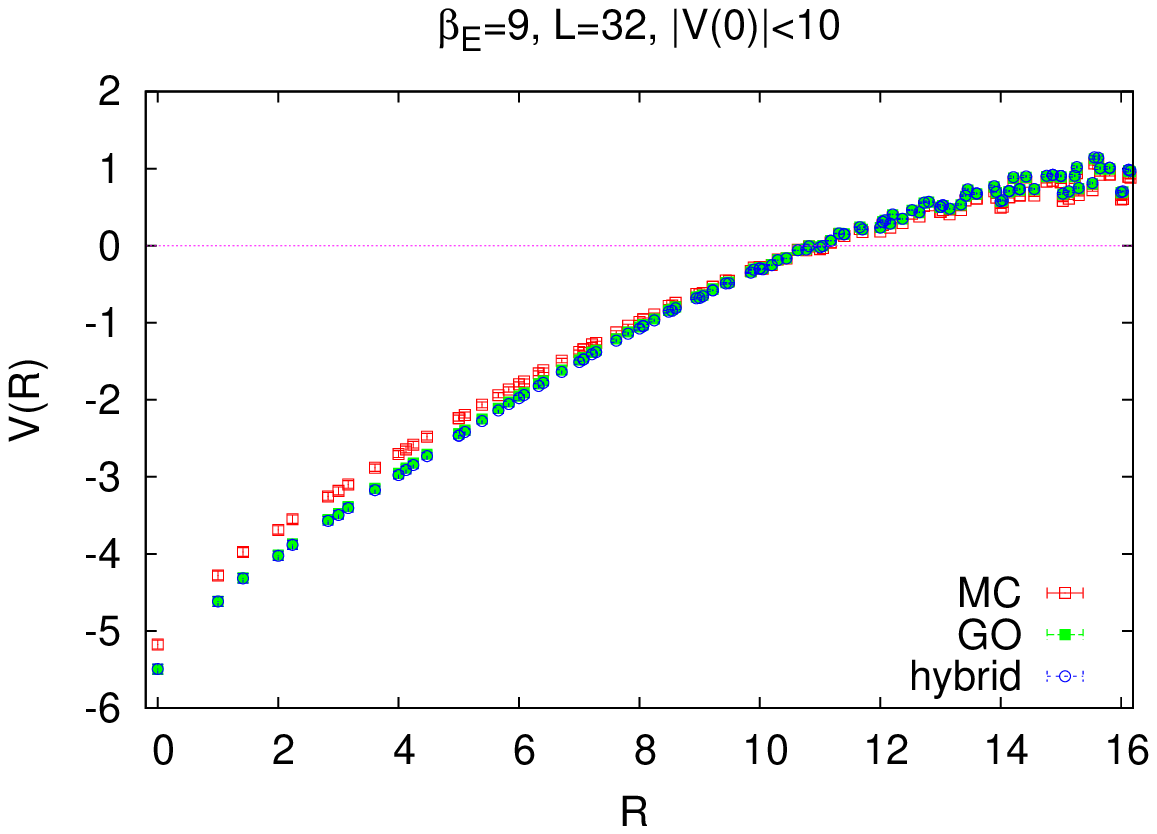}} \\
      \resizebox{70mm}{!}{\includegraphics{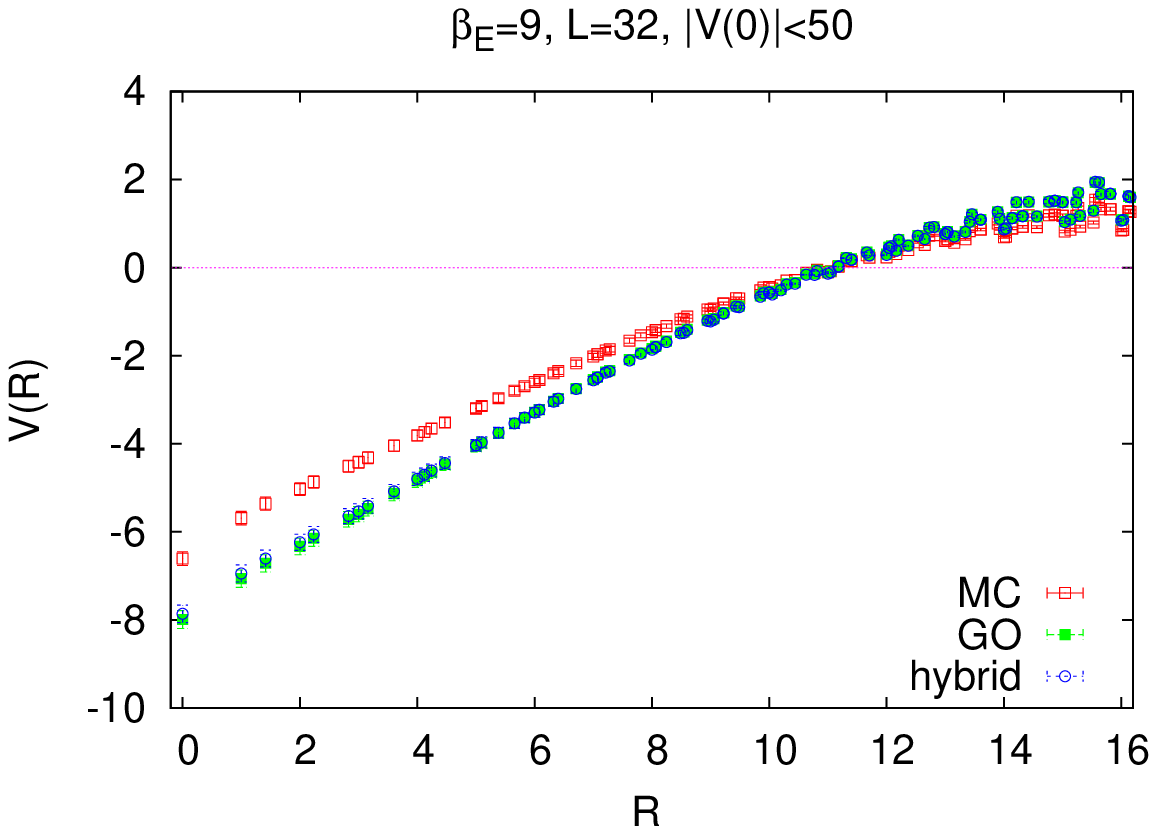}} &
      \resizebox{70mm}{!}{\includegraphics{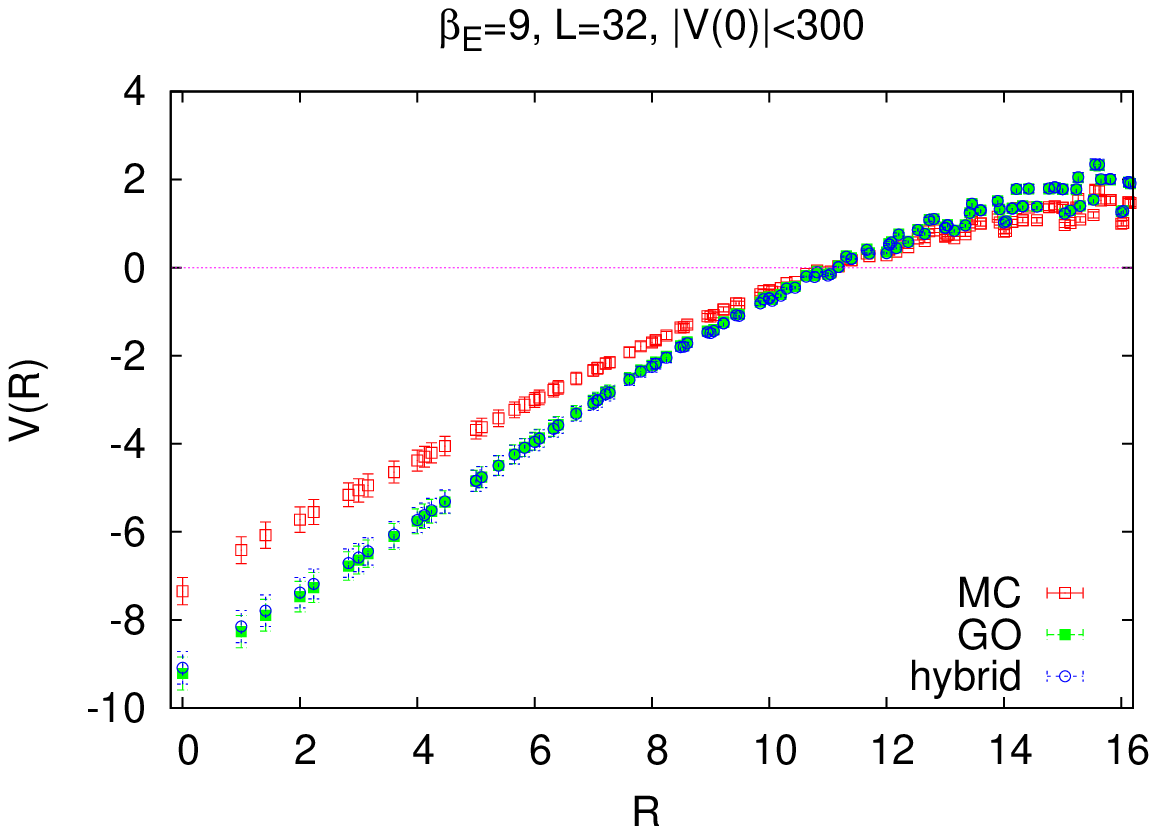}} \\
    \end{tabular}
    \caption{Data for the Coulomb potential at $\b_E=9$ and $L=32$, derived from
    MC, GO and hybrid simulations, with a cut on the data, discarding configurations for
    which $|V_0|$ is greater than 5, 10, 50, and 300, respectively.}
    \label{Vc}
  \end{center}
\end{figure*}

    Figure \ref{GOghost} shows the equal-times ghost propagator $G(R)$ computed in a standard Monte Carlo simulation on a $32^3$ lattice at $\b_E=9$.  On the same plot we see the corresponding results obtained by generating lattices
with probability distribution $\Psi_{GO}^2$ and $\Psi_{hybrid}^2$ by the methods of \cite{Greensite:2007ij}, transforming to Coulomb gauge, and evaluating the ghost propagator, in each case using the appropriate values of $\b,m_L$ corresponding to $\b_E=9$.  It can be seen that the agreement between Monte Carlo, GO, and hybrid results is almost perfect.  

    The agreement  for the Coulomb potential $V_c(R)$ is not as good. In Fig.\ \ref{Vc} we display the data from MC, GO, and hybrid simulations, again at $\b_E=9$, with a cut in the data, discarding configurations with $|V(0)|$ greater than some bound equal to $5,10,50,300$.    If we restrict the data set to configurations with $|V(0)|<5$, then the agreement between MC, GO, and hybrid results is again almost perfect.  Roughly half of all configurations meet this criterion.  The agreement is still fairly good for $|V(0)|<10$, which accounts for about 80\% of all configurations.  However, as the cut is gradually removed, the Coulomb potential derived from GO and hybrid simulations, while roughly linear in $R$, deviates quantitatively from the MC result.  But how can there be such a noticeable deviation when the ghost propagators agree so accurately, without any cuts at all?  The explanation probably has to do with a discrepancy in the tail of the probability distribution.  If two probability distributions agree in their lower moments, but disagree in higher moments, then it means that the two distributions agree pretty well where the probability is substantial, but disagree in the tail of the distributions.  That is what seems to be going on here.

      What was found already in ref.\  \cite{Greensite:2010tm} is that the Coulomb potential is quite sensitive to a comparatively small number of ``exceptional" configurations, in which the lowest eigenvalue of the Faddeev-Popov operator $-\nabla \cdot D$ is far below the average value for the lowest eigenvalue.  The reason that such exceptional configurations are relevant for the Coulomb potential, but not the ghost propagator, is presumably because the ghost propagator involves only one factor of the inverse {F-P} operator, while the Coulomb potential involves two factors.  Because the inverse F-P operator becomes singular as the lowest eigenvalue $\l_0$ approaches zero,  higher powers of the inverse F-P  operator (such as the Coulomb potential) will be more sensitive to infrequent configurations with exceptionally low values of $\l_0$ than lower powers (such as the ghost propagator).  The probability distribution of infrequent configurations is, of course, governed by the tail of the probability distribution.  So our interpretation of the ghost and Coulomb propagator results is that $\Psi_{GO}^2$ and $\Psi^2_{hybrid}$ agree quite closely with each other, and with the probability distribution of the true Yang-Mills vacuum wavefunctional $\Psi_0^2$, in the ``bulk" of the distribution.  The Coulomb potential data suggests, however, there is some small disagreement in the tail of the distribution.

      In general, our results for the Coulomb gauge ghost propagator and Coulomb potential with the new fitting procedure for $\b,m$ agree quite closely with our previous results (based on setting $\b=\b_E$) reported in ref.\ \cite{Greensite:2010tm} (for a quantitative comparison, cf.\ \cite{Greensite:2010yp}). The GO and hybrid results are, once again, virtually indistinguishable.  Since both choices of parameters, and the GO and hybrid wavefunctionals, have about the same dimensional reduction limit, our results suggest that the quantities we have computed, at the couplings we have employed, are mainly sensitive to that limit.  \\

\section{\label{conclusions}Conclusions}

     We have compared several suggestions for the Yang-Mills vacuum wavefunctional  to the true Yang-Mills vacuum wavefunctional in 2+1 dimensions, whose exact form is unknown, but whose relative magnitudes in any set of lattice configurations can be obtained numerically.  Three types of lattice configurations were studied:  abelian plane wave configurations, non-abelian constant configurations of fixed amplitude but varying ``non-abelianicity," and non-abelian constant configurations of maximal abelianicity but of differing wavelengths and varying amplitudes.  For purposes of comparison, the physical scale was set by taking the string tension to be $\sqrt{\s}=440$ MeV.

      For abelian plane waves, up to the shortest wavelength corresponding to $p^2=2.5$ GeV${}^2$ that we have investigated, the GO and Karabali-Kim-Nair proposals are almost indistinguishable, and both agree very well with the values obtained for the true vacuum wavefunctional, evaluated on these configurations.  The Coulomb gauge wavefunctional can also fit the plane wave data with an appropriate choice of parameters, providing in particular that the renormalization constant $c_1$ in eq.\ \rf{gapeq} is set equal to zero.   Both the GO and KKN wavefunctionals  reduce to the dimensional reduction form $\exp[-\m \int F^2]$ at long wavelengths, and it seems likely that this is also true for
the Coulomb gauge proposal, in this special case of abelian configurations, for the choice of renormalization constant $c_1=0$.

  For non-abelian configurations, we have suggested a gauge-invariant wavefunctional which reduces to the KKN proposal for abelian configurations, and incorporates the covariant Laplacian and eigenvalue subtraction of the GO proposal, which we have termed the ``hybrid" wavefunctional.  Both the GO and hybrid wavefunctionals have the dimensional reduction form when restricted to configurations which, when expanded in eigenstates of the covariant Laplacian, contain only low-lying eigenmodes.  Once again, the GO and hybrid wavefunctionals are almost indistinguishable when evaluated on non-abelian constant configurations, and this is probably because they have almost the same dimensional reduction limit.  We find that the GO and hybrid wavefunctionals are in good agreement with the true vacuum wavefunctional for non-abelian constant configurations, as well as for abelian plane waves. The Coulomb gauge wavefunctional, however, which does not have the dimensional reduction property for non-abelian lattices, does not seem compatible with the data for non-abelian constant configurations, particularly the data with variable non-abelianicity.

       The Coulomb gauge wavefunctional has been used to compute Coulomb gauge ghost and gluon propagators, with results in 2+1 dimensions, reported in \cite{Feuchter:2007mq}, indicating a Coulomb potential rising almost (but not quite) linearly.  We have also computed these quantities by direct simulation of the GO and hybrid wavefunctionals.  The GO and hybrid results agree with one another, and almost perfectly with the lattice Monte Carlo results for the ghost propagator.  The GO and hybrid wavefunctionals also lead to an apparently linear Coulomb potential and agree very closely with each other.  On the other hand there is some difference in the GO and hybrid Coulomb potentials in comparison to the lattice Monte Carlo results, and this can be attributed to a difference associated with exceptional configurations with unusually small values of the lowest Faddeev-Popov eigenvalue.  Thus the GO and hybrid  wavefunctionals would seem to agree with the true Yang-Mills vacuum wavefunctional for the bulk of the probability distribution, but there would appear to be a small disagreement in the tail of the distribution.

     The main effort in this article has been to calculate the relative magnitudes of the true vacuum wavefunctional
on particular sets of lattice configurations; namely, abelian plane waves and non-abelian constant configurations, and to compare those results with a number of proposals for the vacuum state.  We have found that the lattice data for the abelian plane waves have been nicely reproduced by all proposals considered, while good agreement with the data for non-abelian constant configurations appears to require wavefunctionals with the property of dimensional reduction.

\acknowledgments{J.G.'s research is supported in part by
the U.S.\ Department of Energy under Grant No.\ DE-FG03-92ER40711.
A.P.S's research is supported in part by the  US Department of Energy grant under
contract DE-FG0287ER40365. M.Q.~and H.R.~are supported by DFG under
contract DFG-Re~856/6-3.  \v{S}.O. is supported in part by the Slovak Grant Agency for Science, Project VEGA No.\ 2/0070/09, by ERDF OP R\&D, Project CE QUTE ITMS~26240120009, and via CE SAS QUTE. }

\appendix*

\section{Numerical details}

\begin{table*}[htb]
\begin{center}
\begin{tabular}{|c|c|c|c|c|c|} \hline
   $\b_E$   &    $L=16$  &  $L=24$  &  $L=32$   &  $L=40$   &  $L=48$    \\  \hline
            6         &    (0,0.5)    &    (0,1.0)       & (20,1.5)    &   (30,2.5) &  (60,3.5)  \\
      9         &    (3, 0.25) &  (5, 0.5)   & (50,0.7)    &  (10,1.3)  &   (20,1.8)  \\
    12         &    (2,0.17)  &  (7, 0.28) & (12,0.53)  &  (20,0.75) & (30,1.0)   \\  \hline
\end{tabular}
\end{center}
\caption{Values of $\a,\g$ used in eq.\ \rf{awave} to generate abelian plane wave configurations with wavelength $\l=L$ equal to the lattice extension, and $\b_E=6,9,12$.}
\label{Ag-ab}
\end{table*}

\begin{table*}[htb]
\begin{center}
\begin{tabular}{|c|c|} \hline
   $\b_E$      &  \{ ($\alpha,~\gamma$) \}     \\  \hline
      6         &    (2,0.15) , (15, 0.20) , (32,0.20) , (60,0.22) , (86,0.24) , (107, 0.26)   \\
      9         &    (2,0.09) , (10, 0.10) , (25,0.13) , (50,0.14)    \\
    12         &    (1.3,0.06) , (4, 0.06) , (10,0.065) , (20,0.08) , (27,0.083) , (35,0.083)    \\  \hline
\end{tabular}
\end{center}
\caption{Values of $\a,\g$ used in eq.\ \rf{nac1} to generate non-abelian constant configurations with maximal non-abelianicity, on a $32^2$ lattice and $\b_E=6,9,12$.}
\label{Ag-nab}
\end{table*}

    Evaluation of $R_{GO}[U]$ involves dealing with a kernel
\beq
          K^{ab}_{xy} = \left({1 \over \sqrt{-D^2 - \l_0 + m^2}} \right)^{ab}_{xy}
\eeq
which, on a lattice of extension $L$, calls for inverting the square root of a $3L^2 \times 3L^2$ matrix.
The numerical evaluation in this case can be accelerated using the Zolotarev approximation, for which
\beq
          {1\over \sqrt{X}} \approx  a_1 \mathbbm{1} + {a_2 \over X + b_2 \mathbbm{1}}
          + {a_3 \over X + b_3\mathbbm{1}} + {a_4 \over X + b_4 \mathbbm{1}} \; ,
\label{zolo}
\eeq
where $X$ is a matrix, and the coefficients are given by \cite{Kennedy:2004tj}
\beq
a1 &=& 0.3904603901
\non \\
a2 &=& 0.0511093775
\non \\
a3 &=& 0.1408286237
\non \\
a4 &=& 0.5964845033
\non \\
b2 &=& 0.0012779193
\non \\
b3 &=& 0.0286165446
\non \\
b4 &=& 0.4105999719 \; .
\eeq
In fact, what one really wants is the vector
\beq
            u^a_x = K_{xy}^{ab} F_{12}^b(y) \; ,
\eeq
and we found it convenient to compute this vector numerically using the Matlab software package.  In Matlab, computation of the vector $\vec{u} = M^{-1} \vec{w}$, given the matrix $M$, requires only a single line of code:
$u = M \backslash w$.  One first defines
$X = -D^2 - \l_0 \mathbbm{1} + m^2 \mathbbm{1}$ to be a sparse matrix, and then sets $Y_2 = X + b_2 \mathbbm{1}$ etc.  The vector $\vec{u}$ with components $u^a_x$ is then obtained by the Matlab statement
\beq
          u = a_1*\mathbbm{1} + a_2*(Y_2\backslash F) + a_3*(Y_3\backslash F) + a_4*(Y_4\backslash F) \; ,
\non \\
\eeq
and we finally take the inner product
\beq
           R = {\b \over 4}   F_{12}^a(x)  u^a_x \; ,
\eeq
with an implicit summation over lattice sites $x$ and color indices $a$.
All the matrix operations, including the determination of $\l_0$, can be carried out numerically using sparse matrix techniques, which results in a considerable savings in computation time, often by an order of magnitude or more in our calculations.  We have checked the accuracy of the Zolotarev approximation by evaluating $R$ numerically, in several cases, without this approximation, and have found the results with and without the approximation to differ only at the third significant digit.  This is sufficient for our purposes.  In the case of $R_{hybrid}$ the formula \rf{zolo} is not directly applicable, and the numerical evaluation was carried out without the help of the Zolotarev approximation.

    In the Monte Carlo simulations, we set up eight runs each time with the same parameters, but different seeds for the random number generator.   Each run is itself a number of independent jobs, which we refer to as ``cycles", whose results for $-\log(N_n/N_T)$ are averaged together at the end of the run.  At the beginning of each cycle the links are all set to the identity matrix, except for the spacelike links on the $t=0$ plane, which are set to the first ($n=1$) configuration out of the set of $\{U_i^{(n)}(x,t=0)\}$ of non-abelian constant configurations.  The lattice at $t\ne 0$ then thermalizes for 5000 sweeps with the $n=1$ configuration at $t=0$ held fixed.   All timelike links are fixed to the unit matrix, except for the timelike links at $t=L/2$, which are updated in the usual way.   After thermalization we carry out another 30000 sweeps, with the configuration at $t=0$ updated only once every 40 sweeps.   On reaching the $t=0$ plane every 40th sweep, we carry out 20 Metropolis ``hits";  i.e.\ the Metropolis algorithm is used to update the $t=0$ plane, and at each hit the plane is changed to a new configuration (or not, depending on the result of the algorithm), and the appropriate configuration counter $N_n$ is incremented.  At the end of each cycle the value for $-\log(N_n/N_T)$  for each configuration $n$ is recorded.  At the upper end (higher $n$) it is usually the case that $N_n=0$ on one or more cycles; all data from these higher $n$ configurations are deemed statistically unreliable, and discarded.   The number of cycles (used for eight runs at the same set of parameters) varied from a minimum of 17 to a maximum of 70, but was mostly around 30.   The result for the slope of $-\log(N_n/N_T)$ vs.\ $R[U^{(n)}]$ was obtained from the best fit to the data in each run, and the results from the eight independent runs were used to estimate the error.

     Finally we record, in Tables \ref{Ag-ab} and \ref{Ag-nab}, the values of $\a,\g$ used, in eqs.\ \rf{awave} and \rf{nac1}, to generate sets of abelian plane waves and non-abelian constant configurations with varying amplitudes.  The aim, in choosing parameters, was to keep
the variation of $r_n=-\log(N_n/N_{tot})$ in a relatively small range $\Delta r_n \approx 4$ (otherwise, because of the exponential falloff, there would be few or no data points at the larger values of $n$).  In the case of non-abelian constant configurations, we choose different $\a$ values so as to sample the slope of $-\log(N_n/N_{tot})$ vs. $R[U]$ in a small interval of $\D R$, centered around a wide range of values of $R$, as explained in subsection \ref{D}.

\bibliography{vac-version-7-5}
\end{document}